\documentclass[journal]{IEEEtran}
\usepackage[utf8]{inputenc}
\usepackage[T1]{fontenc}
\usepackage{amsmath,amssymb}
\usepackage{booktabs}
\usepackage{tabularx}
\newcolumntype{Y}{>{\centering\arraybackslash}X}
\newcolumntype{L}{>{\raggedright\arraybackslash}X}
\usepackage{url}
\usepackage{cite}
\usepackage{microtype}
\begin{document}

\title{Tool Waiting and Re-arrival in Compile-Time-Static LLM Serving: Cost Mechanisms and Configuration Selection}

\author{Dongkyeom~Jang, In-Nea~Wang, and Junho~Jeong%
\thanks{Dongkyeom Jang and Junho Jeong are with the Department of Computer Science and Artificial Intelligence, Dongguk University, Seoul, Republic of Korea (e-mail: jangelliot0404@dgu.ac.kr; yanyenli@dongguk.edu). Junho Jeong is the corresponding author.}%
\thanks{In-Nea Wang is with the Convergence College of IoT Innovation, Dongguk University, Seoul, Republic of Korea (e-mail: innea@dgu.ac.kr).}}

\markboth{}%
{Jang \MakeLowercase{\textit{et al.}}: Tool Waiting and Re-arrival in Compile-Time-Static LLM Serving}

\maketitle

\begin{abstract}
In agentic LLM services, a session calls an external tool, waits for it, and re-arrives to continue inference. Statically compiled NPU serving can fix the batch bucket set, the maximum batch size, and the number of KV cache slots at compile time. We define such an environment as a compile-time-static serving substrate and analyze the execution-time cost that tool waiting and re-arrival incur in it. On a single LLM instance, we run synthetic workloads following a measured tool waiting time distribution and compare, on the same inputs, a baseline configuration with settings \{1, 2, 4, 8\}, 8, and 8 against configurations that change some of them. Because re-arrival times differ across configurations, we build a simulator that replays request processing in time order, select the candidate with the lowest predicted cost among 2,077 configurations, and validate it on new inputs. We identify three mechanisms: discrete batch alignment, KV cache survival, and prefill interference. At a concurrency of 6, absent from the bucket set, tool waiting lowered the padding ratio (0.235 to 0.120) yet increased decode execution time 1.51-fold, so padding alone did not indicate cost. On new inputs at a concurrency of 8, where the baseline reused KV in 9 of 24 re-arrivals, enlarging the maximum batch size alone cut execution cost by 8.25\%, and the selected configuration, which also adjusted the bucket set, by 9.72\%. Where 17 of 18 re-arrivals were already reused, the effect was 0.59\%. Compile-time configurations should thus be selected by diagnosing KV reuse loss and the resulting change in execution.

\end{abstract}


\section{Introduction}\label{sec:1}

LLM services are expanding beyond question answering toward agentic operation, in which the model calls external tools and continues reasoning from their results. A session then repeatedly passes through waiting intervals in which it waits for a tool call---web search, code execution, an external API---to finish. During such an interval, the session's KV cache remains in accelerator memory, occupying memory without being used for computation. If that KV is evicted while other requests are admitted, recomputation becomes necessary when the tool call completes and inference resumes. How this idle memory occupancy and recomputation appear as actual system cost depends on the execution and resource management structure of the accelerator.

As AI investment and data center construction grow, the accelerators used for LLM serving are diversifying from GPUs to NPUs and other inference accelerators. Many NPU serving stacks do not handle dynamic shapes directly; instead, they compile static-shape execution graphs and, at run time, fit requests to the prepared execution forms through bucketing or padding. The vLLM serving paths for AWS Neuron \cite{ref1}, Google TPU \cite{ref2}, Intel Gaudi \cite{ref3}, and Huawei Ascend \cite{ref4} all follow this structure. In an environment where, in addition, the batch bucket set, the maximum batch size, and the number of KV cache slots are all fixed at compile time, requests must be processed within the execution forms and the number of KV slots set at compile time, even as the number of active requests, or the number of sessions whose KV must be retained during tool waiting, changes during execution. This paper calls such an environment a compile-time-static serving substrate and takes it as the object of study.

Prior work related to idle occupancy of accelerator memory falls into KV management policies and analyses of static execution structures. In GPU environments, methods have been proposed for managing KV caches that become idle during tool waiting: predicting tool duration to keep the cache in memory \cite{ref5}, moving the waiting KV cache to other memory \cite{ref6}, and controlling eviction with awareness of the agent execution structure \cite{ref7,ref8}. These approaches reduce recomputation and memory occupancy by adjusting run-time decisions about retaining, evicting, and moving KV, and their applicability and effect depend on the KV management functions and the allocation and eviction units provided by the target stack. Separately, the characteristics of NPU and static-graph LLM serving have been analyzed: nonlinear latency across batch buckets \cite{ref9}, cases in which the fixed-shape constraints of static-graph deployment restrict KV management design \cite{ref10}, and serving strategies on multi-core NPUs reviewed systematically \cite{ref11}. These studies, however, mainly target conventional request serving; the intersection between the agentic arrival process, in which existing sessions rejoin the execution flow after finishing tool calls, and the compile-time-static serving substrate has not been treated systematically.

This gap is more than the absence of a combination of two research areas. In agentic workloads, re-arrivals occur repeatedly as existing sessions return to a runnable state after tool calls, in addition to user requests. Because tool durations differ across sessions, re-arrivals are distributed irregularly along the time axis, and the number of runnable sessions at each moment keeps changing. In a compile-time-static serving substrate, the batch bucket and KV slot options that could respond to this variation are constrained in advance at compile time. The time-varying agentic arrival process thus collides directly with the resource structure fixed beforehand. Applying existing tool-aware KV management techniques in this environment requires distinguishing the retention and eviction policies that can be adjusted at run time from the number of KV slots and the bucket configuration fixed at compile time. This paper focuses on the latter constraint and identifies what costs tool waiting and re-arrival create under a fixed resource configuration, and through which execution mechanisms those costs arise.

To this end, we ran synthetic agentic workloads that include tool waiting and re-arrival on a single language-model instance deployed across four devices of a Rebellions CA25 NPU server, and analyzed, through measurement and intervention experiments, the paths by which the re-arrival process affects resource usage and execution cost. The paper makes four contributions.

First, we identify from measurements three mechanisms---discrete batch alignment, KV-cache survival, and prefill interference---through which the re-arrival process affects device time and the delay of other sessions. In particular, by adding one bucket, recompiling, and performing a paired comparison with the same input plan, we confirm that the bucket grid is the cause of the padding difference.

Second, we show the structure by which the three mechanisms affect one another in the same arrival process. An intervention that changes the number of KV slots confirms that resolving reuse failures reduces the recomputation work of re-arrivals, and a separate controlled experiment supports the possibility that the additional prefill leads to delays in background sessions. Within the same paired comparison, we also show that there is a conflicting regime in which tool waiting improves batch alignment while worsening KV reuse.

Third, we evaluate how accurately a simulator that reflects the main execution rules of the three mechanisms predicts cost ratios between configurations within the validated range, and report the size and direction of the unexplained residual. Because the residual contains approximation error in the cost model and measurement and execution variation as well as unmodeled mechanisms, we do not interpret its size as the size of any particular factor.

Fourth, we apply the identified mechanism structure to configuration selection. We explore 2,077 candidate configurations with the validated simulator and evaluate the selected configurations by measurement on new inputs. We show that, where the baseline configuration loses substantial reuse, changing only the compile parameter \texttt{batch\_size}, which sets both the number of KV slots and the execution limit, captures most of the gain of the improved configuration; that where reuse is already retained, the effect of adding slots is limited; and we report the point at which the gain from additional slots stops. Based on this, we present a configuration selection procedure that diagnoses the losses in a workload, explores candidates, and validates on new inputs.

The paper is organized as follows. Section~\ref{sec:2} describes the execution structure of the target NPU serving environment and the measurement methods. Section~\ref{sec:3} presents the measurement and intervention results for each of the three mechanisms, with the level of confirmation distinguished, and then summarizes how the three mechanisms couple through the arrival process. Section~\ref{sec:4} constructs a simulator that reproduces execution in event order and evaluates its prediction accuracy and residual. Section~\ref{sec:5} evaluates by measurement the gain of configuration selection and the conditions under which it appears, and Section~\ref{sec:6} discusses limitations and concludes. Details of the decision criteria, the recompilation check, the attribution of eviction events, and the appendix computations are provided as supplementary material.

\section{Methods}\label{sec:2}

\subsection{Target System}\label{sec:2-1}

The experiments used a Rebellions RBLN-CA25 server. The server holds 8 cards, is recognized as 32 devices in total, and has 15.7 GiB of memory per device. We deployed one language-model instance on 4 of these devices and left the remaining devices unused. All measurements reported below target this single instance.

The serving software was vLLM 0.22.0+cpu, vllm-rbln 0.11.1, optimum-rbln 0.11.1, and rebel-compiler 0.11.1.post1, with kernel driver KMD version 3.2.2. The language model was Qwen3-4B in bfloat16 precision. \texttt{max\_seq\_len} was set to 8,192 and \texttt{num\_devices} to 4. The attention implementation used the \texttt{eager} path, the default when \texttt{kvcache\_partition\_len} is not specified \cite{ref12}. The KV cache allocation and eviction structure described below refers to this path.

The model has 36 layers, 8 KV heads, and a head dimension of 128. Since K and V are each stored in bfloat16, the KV capacity per token is

\begin{equation}
36\times8\times128\times2\times2=147{,}456\;\mathrm{B}=144\;\mathrm{KiB}
\end{equation}

\subsection{Execution Structure}\label{sec:2-2}

Decode runs in steps, and the batch size of each step is chosen from the compiled bucket set \texttt{decoder\_batch\_sizes}. Let $n_t$ be the number of requests actually batched in step $t$ and $\mathcal B$ the compiled bucket set. The selected bucket $b_t$ is

\begin{equation}
\begin{gathered}
b_t=\min\{b\in\mathcal B\mid b\ge n_t\},\\
1\le n_t\le\texttt{batch\_size}
\end{gathered}
\end{equation}

In a step with $b_t>n_t$, only $n_t$ of the $b_t$ batch positions hold actual requests, and the remaining $b_t-n_t$ positions are filled with padding. We confirmed this correspondence by exhaustively cross-checking the source code against measured logs. The baseline configuration uses $\mathcal B=\{1,2,4,8\}$ and \texttt{batch\_size=8}. If \texttt{decoder\_batch\_sizes} is not specified, only a single bucket equal to \texttt{batch\_size} is generated.

The KV cache is managed in blocks; the size of a block (\texttt{kvcache\_block\_size}) and the total number of blocks (\texttt{kvcache\_num\_blocks}) are fixed at compile time. On the eager path, these two values are set to \texttt{max\_seq\_len} and \texttt{batch\_size}, respectively, which we confirmed in all 7 configurations of the same language model compiled under different conditions. Because the block size equals the maximum sequence length, one block can hold the KV of one sequence. We call such a storage space a KV slot. Each slot reserves a fixed amount of memory for the KV cache of a sequence at the maximum length. We therefore express KV capacity in terms of the number of slots hereafter. The baseline configuration has 8 KV slots. KV eviction uses FIFO, and re-access does not refresh eviction order.

The scheduler does not place prefill and decode in the same execution step. Section~\ref{sec:3} analyzes how the bucket selection, KV storage and eviction, and exclusive execution structure affect tool waiting and re-arrival.

\subsection{Workload and Comparison Design}\label{sec:2-3}

Each session starts with a prompt of 800--1,600 tokens and generates 32--256 tokens. After completing the first request, the session waits for a predetermined tool waiting time and then sends a follow-up request. The follow-up request appends 8 tokens of additional context to the previous prompt and the generated text, and generation proceeds again after re-arrival. If the prefix of the previous prompt is preserved and its KV remains, that part is reused and only the remainder is computed; if the KV has been evicted, the entire re-arrival prompt is recomputed. Each session contains exactly one tool waiting interval. Multiple tool calls would introduce both the effect of prompts lengthening as context accumulates and the effect of repeated calls; excluding these isolates the effect of re-arrival itself.

We use synthetic prompts, and the input plan sets the prompt length and the generation-length limit of each request. The same input plan is applied to the compared conditions. In the stored response records, the number of tokens actually generated matched the planned limit in every run compared in the main text, and no request finished before reaching the limit. Generation uses greedy decoding at temperature 0 with a fixed seed. Cache reuse requires both survival of the target KV and preservation of the token IDs of the reused prefix, but the stored data cannot exhaustively verify byte-level identity of generated text across conditions or preservation of prefix token IDs. We therefore separately confirmed that prompt token counts matched across conditions and measured the actual number of cached tokens; the range in which reuse actually occurred is reported from measurements in Section~\ref{sec:3-2}.

The control condition uses the same input plan with only the tool waiting interval removed. Each session still consists of 2 requests, and the re-arrival request is constructed and sent immediately after the response to the first request is fully received; the measured time from reception to transmission is 0.5--1.2 ms. The request boundaries, the token count of the input prompt, the planned number of generated tokens, and the construction rule of the re-arrival prompt are identical in the two conditions; the only difference is the presence or absence of the tool waiting interval. The actual prefill computation may differ between the two conditions depending on cache survival, and this difference is what we observe.

The tool waiting time distribution was constructed from TraceLab \cite{ref13} v0.0.2, which covers usage records of Claude Code and Codex by 52 users from September 2025 to July 2026, comprising 743,819 tool calls in 8,058 sessions. Excluding 4 tools that wait for a human response, we sample a tool from the remaining 43 with probability proportional to its call frequency, and draw its latency from an inverse cumulative distribution constructed to pass exactly through the quantiles reported by TraceLab. We use the executor-reported internal latency when available and the wall-clock difference otherwise. Latencies were capped at 60 seconds; the cap applied to 2.51\% of samples.

As the number of concurrent sessions N increases, both the number of active requests in each step and the number of other sessions' requests arriving during one session's waiting interval increase on average. These two quantities determine bucket selection and KV slot pressure, respectively, so N is the main control variable. The tested values of N were \{3, 4, 5, 6, 7, 8, 10, 12, 16\}. Here 8 equals the \texttt{batch\_size} of the baseline configuration and 16 is twice that value, corresponding to the case in which the number of concurrent sessions equals the number of slots and the case in which it exceeds the number of slots and queueing occurs. The remaining values were chosen so that the padding ratios against the bucket set \{1, 2, 4, 8\} differ. Of these, \{6, 8\} form the confirmatory range and \{10\} the exploratory range, because the validation range of the simulator in Section~\ref{sec:4} corresponds to the confirmatory range.

Measurement begins with three input plans generated with different random seeds, each called a replicate. Within one replicate, all conditions receive exactly the same input plan, so comparisons between conditions within a replicate control for differences in the input plan. Across replicates, differences in the input plan and variation in execution itself appear together; the size of run-to-run variation for the same input is gauged with separate no-treatment repeated measurements (Supplementary Material S1). The input plan fixes each session's prompt length and text, generation length, tool waiting time, and session composition. Sessions were designed to start simultaneously, but actual first transmission times were spread over 2--12 ms, and the actual time of re-arrival is set by adding the waiting time to the completion time of the first request and thus varies with the execution outcome. The input plan fixes the inputs to execution but not the times that execution produces. The total number of measurements for one value of N is the number of replicates (3) multiplied by the number of compared conditions.

\subsection{Measurement Metrics and Experimental Procedure}\label{sec:2-4}

The serving stack does not expose the pair of the actual request count and the selected bucket per step, so we applied an observation-only patch to the decoder preprocessing function of \texttt{vllm-{}rbln}, where that decision is made. The patch adds 5 lines of code including one log statement and does not touch control flow, bucket selection, or KV allocation; the SHA256 of the target file was fixed and recorded in every measurement record. Actual KV reuse was determined from \texttt{prompt\_tokens\_cached\_total}. \texttt{prefix\_cache\_hits\_total} was excluded because it was found to report false positives, reporting hits even after the corresponding KV slot had been evicted.

In this study, device time is defined as the sum of time attributed to execution activity. It sums the measured decode step cost and the server-side prefill elapsed time and thus does not denote pure accelerator occupancy alone. Wall-clock time from the start to the end of a run is not used as a cost metric because it can include periods in which no session's execution is progressing during tool waiting, and utilization is not used as an efficiency metric because it includes time spent on padding computation and recomputation. We use device time as the cost metric. Because padding and recomputation lengthen execution and thus device time, they can be read as additional cost, and cost comparisons are made only between runs that process the same input plan.

Because the serving stack does not report device time directly, we cross-checked two channels. Channel A applies the measured decode step cost to the $(n_t,b_t)$ pairs of the step sequence and the prefill cost model to the number of prefill tokens actually computed; it is a reconstruction that depends on the two cost models. Channel B is the union of in-flight intervals taken from the request transmission and completion times recorded by the client; it does not depend on the cost models but includes queueing, HTTP, and scheduler overhead, so it is an activity-time proxy rather than device time itself. For the summed values of the three replicates of the baseline condition, the channel difference r = B $-$ A was about 1.0--1.7 s; because this may vary with the input and execution condition, it is not treated as a fixed cost independent of N.

A decision is made only when the two channels satisfy a predefined agreement criterion. The compared quantities are the ratios relative to the baseline condition computed by each channel, using values summed over the three replicates. In the initial configuration cost validation, a fixed tolerance of 0.02 was applied to the difference between the two channels' cost ratios. For the subsequent re-measurement with new inputs, the tolerance rule was changed to $\tau$ = max(0.02, $r_{BASE}$/$B_{BASE}$) and registered before re-measurement, where $B_{BASE}$ is the channel B value summed over the three baseline replicates and $r_{BASE}$ is $B_{BASE}$ $-$ $A_{BASE}$ from the same data. This rule is an empirical criterion for the discrepancy allowed between the two channels, not a rigorous bound derived from overhead, and the default 0.02 is a predetermined decision criterion, not a value derived statistically from measurement variation. A sensitivity analysis of the default and no-treatment repeated measurements of whether the tolerance covers run-to-run variation are given in Supplementary Material S1. The withheld decisions on the initial data are retained, and the applicable targets and results of the two criteria are reported separately in Section~\ref{sec:4}.

The server was restarted for every condition so that cache state would not carry over, and the execution order of conditions within a replicate was changed from replicate to replicate so that systematic differences from running earlier or later would not concentrate on a particular condition. In intervention experiments, we chose one compile parameter to change and compared two configurations that share the input plan and the remaining specified conditions; we refer to this as a paired comparison. Execution characteristics that change with the parameter are reported separately in each experiment. Predictions and decision criteria were committed to the repository before measurement and not changed afterward, and the order of commit time and measurement time is on record. The seeds of the measurements referenced in constructing the simulator differ from those of the validation measurements, which used new seeds not used in any earlier measurement.

\section{Three Mechanisms by Which Re-arrival Affects Execution Cost and Concurrent Session Delay}\label{sec:3}

This section analyzes discrete batch alignment, KV cache survival, and prefill interference, which correspond to the execution structure in Section~\ref{sec:2-2}. Sections~\ref{sec:3-1} and~\ref{sec:3-2} address the time cost of decode and prefill work, and Section~\ref{sec:3-3} addresses the waiting that exclusive execution of prefill imposes on background sessions; cumulative waiting time is computed separately from device time. Section~\ref{sec:3-4} summarizes the confirmed range of the three mechanisms and the structure by which they couple through the same arrival process.

\subsection{Discrete Batch Alignment}\label{sec:3-1}

Discrete batch alignment refers to the structure in which the actual request count is executed in one of the compiled buckets. Because step cost is set by the bucket rather than by the request count, a change in the distribution of active requests caused by tool waiting changes both the bucket of each step and the number of steps, and thus the cost. Padding is only one consequence of this structure; this section shows that padding and step count can move in opposite directions.

\paragraph{Structure of step cost}
The batch size of each step is chosen from the compiled bucket set. Within the measured range, step cost can be approximated as a fixed cost determined by the selected bucket plus a cost proportional to the actual number of requests, and the latter is small: each additional request adds 0.041 ms on average. This does not mean that splitting a batch into smaller buckets to reduce padding lowers cost. Let $t_{\mathrm{model}}(b,n)$ be the model execution time when $n$ requests are placed in bucket $b$. Measured with each bucket fully filled, the medians were $t_{\mathrm{model}}(1,1)$ = 9.51 ms, $t_{\mathrm{model}}(2,2)$ = 10.05 ms, $t_{\mathrm{model}}(4,4)$ = 10.36 ms, and $t_{\mathrm{model}}(8,8)$ = 12.40 ms. While the bucket grows eightfold, the cost grows only 1.3-fold, and the smallest bucket takes about 77\% of the time of the largest. That is, the cost of one step (about 9.5--12.4 ms) exceeds the difference between adjacent buckets (0.3--2.0 ms), which in turn exceeds the cost of one request (0.041 ms). Dividing the same work into several small batches therefore yields only a limited reduction in per-step time while the number of steps increases, and total cost can grow.

\paragraph{Padding metric}
The padding ratio $p$ is the fraction of all bucket positions not occupied by actual requests.

\begin{equation}
p=\frac{\sum_t(b_t-n_t)}{\sum_t b_t}
\end{equation}

Let $p_A$ and $p_C$ be the padding ratios with and without tool waiting, and define $\Delta p=p_C-p_A$; a positive $\Delta p$ means the condition with tool waiting has the lower padding ratio. With $h(n)$ the number of steps with $n$ active requests and $b(n)$ the smallest bucket accommodating $n$, $p=\sum_n h(n)[b(n)-n]/\sum_n h(n)b(n)$, so the padding ratio is jointly determined by the distribution of active request counts and the bucket set.

\paragraph{Alignment differences by number of concurrent sessions}
The direction of the padding-ratio difference between the two conditions changes with N (Table~\ref{tab:3-1}). At N=6 the padding ratio with tool waiting was 0.120, lower than 0.235 without it, whereas at N=8 the values were 0.163 and 0.090. This reversal is explained by how well the number of active requests fits the bucket grid. The value 6 is not in the grid \{1, 2, 4, 8\}. Without a waiting interval, each re-arrival follows immediately after the first request completes, so a state with 6 active requests persists and 42\% of all steps run 6 requests in bucket 8 with two positions empty. With a waiting interval, sessions leave at different times, so more steps have active counts of 1, 2, or 4 that match the grid, and the fraction of steps executed without padding rose from 39\% to 79\%. This is the opposite of the hypothesis that tool waiting would spread execution out and increase waste. At N=8, by contrast, 8 is in the grid and aligns exactly while all sessions are active, but with tool waiting the completion and re-arrival times differ across sessions, steps with 5--7 runnable sessions became more frequent, and running these in bucket 8 raised the padding ratio to 0.163.

We checked the same regularity at N=3 and N=7. Without tool waiting at N=3, 3 active requests run in bucket 4 (padding 1/4 in those steps), and $\Delta p$ was +0.068 and +0.088 in two measurement batches, repeating the direction at N=6. At N=7, 7 requests run in bucket 8 (padding 1/8), and the summed $\Delta p$ of the two batches was +0.018 and +0.026, but the preregistered decision criterion ($R=(1-p_A)/(1-p_C)$; Supplementary Material S4) did not confirm a difference. Because the direction of the difference depends on how often the active count dwells at each value, this regularity is limited to observations on the measured workload distribution.

\begin{table*}[!t]
\caption{Padding and decode device time with and without tool waiting}
\label{tab:3-1}
\centering\small
\begin{tabular}{lccccc}
\toprule
Grid & N & $p_A$ & $p_C$ & $\Delta p$ & Decode device time ratio (with / without tool waiting) \\
\midrule
\{1,2,4,8\} & 3 & 0.118 & 0.198 & \textbf{+0.080} & \textbf{1.43} \\
\{1,2,4,8\} & 4 & 0.035 & 0.061 & \textbf{+0.026} & \textbf{1.78} \\
\{1,2,4,8\} & 5 & 0.171 & 0.268 & \textbf{+0.098} & \textbf{1.27} \\
\{1,2,4,8\} & 6 & 0.120 & 0.235 & \textbf{+0.115} & \textbf{1.51} \\
\{1,2,4,8\} & 7 & 0.158 & 0.180 & \textbf{+0.022} & \textbf{1.38} \\
\{1,2,4,8\} & 8 & 0.163 & 0.090 & $-$0.072 & 1.39 \\
\{1,2,4,8\} & 10 & 0.152 & 0.069 & $-$0.084 & 1.23 \\
\{1,2,4,8\} & 12 & 0.098 & 0.019 & $-$0.079 & 1.05 \\
\{1,2,4,8\} & 16 & 0.039 & 0.033 & $-$0.005 & 1.07 \\
\bottomrule
\end{tabular}
\end{table*}

\paragraph{Intervention: adding bucket 6}
If the observed padding difference originates from the bucket grid, adding bucket 6 should eliminate the padding improvement from tool waiting at N=6. We added 6 to the bucket set, recompiled with the other settings unchanged, and performed a paired comparison, separate from the measurements of Table~\ref{tab:3-1}, that applies the same input plan to the two grids (Table~\ref{tab:3-2}). At N=6, adding bucket 6 reduced the padding ratio without tool waiting from 0.232 to 0.057, and $\Delta p$ reversed from +0.065 to $-$0.040. All three replicates showed the same sign reversal, and at N=8 $\Delta p$ remained negative for both grids.

We verified from the step sequence that the intervention acted through the intended path. Without tool waiting, the step distribution h(n) was identical for the two grids (1,288 steps in total, with matching counts per request count), so the change in p is attributable solely to which bucket each active count runs in. The rule changes only for 5 and 6 active requests: steps with 6 ran in bucket 6 instead of 8, removing 2 padding positions per step, and steps with 5 changed from 5$\rightarrow$8 to 5$\rightarrow$6, reducing padding from 3 to 1. As a result, the 609 bucket-8 steps in the baseline grid became 0. About 81\% of the reduction came from steps with 6 active requests and about 19\% from steps with 5; the reduction in padding slots and a Shapley decomposition of the padding-ratio change give similar shares (81.1/18.9 and 80.9/19.1; Supplementary Material S5).

\begin{table*}[!t]
\caption{Change in padding at N=6 when bucket 6 is added}
\label{tab:3-2}
\centering\small
\begin{tabularx}{\textwidth}{L Y Y Y}
\toprule
Decode bucket set & With tool waiting: $p_A$ & Without tool waiting: $p_C$ & $\Delta p=p_C-p_A$ \\
\midrule
\{1,2,4,8\} & 0.167 & 0.232 & +0.065 \\
\{1,2,4,6,8\} & 0.098 & 0.057 & $-$0.040 \\
\bottomrule
\end{tabularx}
\end{table*}

\paragraph{Conversion to decode execution cost}
Applying the measured step costs to the same step sequences, the condition with tool waiting used more decode device time in all 9 conditions of Table~\ref{tab:3-1}, with cost ratios of 1.05--1.78, including the five combinations ($\Delta p>0$) in which it had the lower padding ratio. When tool waiting disperses requests into smaller batches, padding may decrease but the number of steps increases, and because the fixed part dominates per-step cost, the cost of additional steps exceeds the gain from reduced padding. At N=6 on the baseline grid, the decode step count with tool waiting was 2,193 summed over three replicates, versus 1,308 without. Because this conversion depends on the cost model, we cross-checked it against channel B, which does not; in all 8 comparison pairs of the grid intervention, the difference between the channel A total cost ratio and the channel B ratio was within the preregistered tolerance (ratio differences 0.0021--0.0261, tolerances 0.036--0.066). The check that the costs of the buckets common to both grids are unchanged across recompilation is given in Supplementary Material S2.

\paragraph{Confirmed range}
The causal effect confirmed by intervention extends to the bucket grid producing the padding difference. That re-arrival changes the number of active requests is an observation, the conversion from padding to device time is a computation based on the cost model, and the cost contributions of padding and step count were not separated quantitatively. The selection rule and the grid \{1, 2, 4, 8\} are characteristics of this stack, but fitting execution to a discrete bucket grid can appear generally in systems that fix shapes at compile time; on GPUs, the same selection rule appears in the source code for vLLM's cudagraph capture size grid. Three conclusions follow. The path by which tool waiting increases cost is quantified from the $(n_t,b_t)$ of the step sequence and the bucket fixed costs, which becomes the alignment component of the simulator in Section~\ref{sec:4}. The bucket grid is a compile parameter and can be adjusted to the workload, but what the adjustment removed was padding at a specific size; on the intervention grid as well, the condition with tool waiting used more device time. And padding-type metrics can point in the opposite direction from cost in a compile-time-static serving substrate.

\subsection{KV Cache Survival}\label{sec:3-2}

\paragraph{Reuse signal and range}
On this stack, the unit in which prefixes are managed differs from the unit in which KV is allocated and evicted. The prefix management unit is 128 tokens under LRU, whereas KV is stored in 8,192-token slots evicted by FIFO. When the reuse target fits in one slot, as here, the survival of that slot determines whether reuse occurs, and the number of reused tokens is limited to 128-token units. Actual reuse was measured from the increment of \texttt{prompt\_tokens\_cached\_total} in the sequential experiments and from \texttt{usage.prompt\_tokens\_details.cached\_tokens} in each response in concurrent execution, with per-response values cross-checked against the KV reuse log linked by request ID to confirm which request each value belongs to.

Let $p_0$ be the prompt token count of the previous request and $p_1$ that of the re-arrival request. For requests with observed reuse, the cached token count matched

\begin{equation}
C=128\left\lfloor\frac{\min(p_0,p_1-1)}{128}\right\rfloor
\end{equation}

This held for all 271 re-arrivals with observed reuse among the 806 in the initial analysis data used to derive the formula, and for all 95 reused re-arrivals in a separate validation measurement not used for the derivation. A formula that widens the reuse range into the previous generated output matched only 38 of the 271 and 24 of the 95, only where the generated token count was small enough that both formulas predict the same value, and the formula above matched in every case where the two diverge. The observed reuse range is therefore the 128-token-aligned portion of the previous prompt, excluding the previous generated output. Of the 168 re-arrivals in the validation measurement, 95 reused the entire range given by the formula and 73 had a cached token count of 0; no partial reuse was observed. We did not determine the cause of the 73 failures case by case, and causes other than eviction may be mixed in, but the disappearance of reuse failures in the intervention with more KV slots on the same input structure supports the interpretation that the number of slots and the eviction structure are involved. Hereafter, survival means that the KV corresponding to the reusable range of the previous prompt is retained until re-arrival.

\paragraph{Sequential experiment controlling the number of background requests}
In concurrent execution, request overlap makes it hard to isolate the effect of the number of subsequent requests. We therefore ran a sequential experiment that processes requests one at a time: one target request, then m background requests, then one re-arrival request that reuses the target prefix. Here m is the number of background requests processed during the interval corresponding to tool waiting and is distinguished from N. Target and background prompts are 2,000 tokens each, and the re-arrival request is 2,008 tokens (the target prompt plus 8 tokens); the previous generated output is not included, the generation limit is 8 tokens with greedy decoding, requests are sent with concurrency 1, the server is restarted per replicate, and background prompts share no prefix with the target or with each other.

In the baseline 8-slot configuration, 1,920 tokens ($p_0$=2,000, $p_1$=2,008 in the formula) were reused with up to 6 background requests, and reuse dropped to 0 from 7. In a re-validation with new seeds and prompts and 3 replicates each for m $\in$ \{5, 6, 7, 8\}, all 6 replicates with m $\in$ \{5, 6\} reused 1,920 tokens and all 6 with m $\in$ \{7, 8\} reused 0; the reuse log and the cached-token metric agreed in all 12, and no partial reuse occurred. We call the maximum number of background requests at which reuse is retained the critical request count. It was observed for this request order and slot configuration and is not interpreted as a threshold on N.

\paragraph{Eviction order and KV survival}
The switch from full reuse to 0 without intermediate values is linked to the structure in which the re-arrival request itself demands a new slot. On this stack, a new slot is allocated even to a re-arrival with a cache hit, and the KV in the existing slot is copied into it as the source. If no free slot exists, the slot with the oldest allocation is evicted by FIFO, and re-access does not refresh the order. With 6 background requests, the target KV was reused first and its slot evicted afterward; with 7, the target KV had already been evicted before the lookup. The latter re-arrival was assigned the same slot ID again, but as a reallocation of evicted space, so survival cannot be judged from a slot ID alone and requires the allocation generation and the order of lookup and eviction. The stopping point was predicted in advance as m=6 or m=7 and was observed at m=7. The observed number of evictions was $\max(0, m-5)$, which matched all 21 runs and exceeds the $\max(0, m-6)$ expected from request allocation alone by one in every run in which eviction occurred; this additional eviction arises on the path by which the scheduler requests a dummy block for padding from the same KV pool, which a separate run recording the call path confirmed directly (Supplementary Material S3). The run-wide eviction count alone cannot, however, determine the survival threshold; what determines survival is whether the eviction of the target KV precedes the reuse lookup.

In concurrent execution, the arrival order of re-arrivals differs from run to run. Sorting the eight re-arrivals at N=8 by client transmission time, in five of the six condition--replicate combinations the mean transmission rank of requests that succeeded in reuse was ahead of that of requests that failed; since transmission order was not manipulated independently, this is an association, not a causal relation. Repeating the same N=8 input plan and configuration 10 times with server restarts, six of the eight re-arrivals succeeded in seven runs and five in three runs. Even with the same input, transmission and completion times shift slightly and change the overlap pattern, so in concurrent execution the KV survival of individual requests cannot be expected to be identical across runs.

\paragraph{Intervention with more KV slots}
If the fixed number of slots is involved in reuse failure, a configuration with more slots should improve reuse on the same input plan. We changed \texttt{batch\_size} from 8 to 16 and recompiled; on the eager path the number of KV slots also increases from 8 to 16. The bucket sets are \{1, 2, 4, 8\} (baseline) and \{1, 2, 4, 8, 16\} (changed). In three paired comparisons at N=8, the maximum actual batch size was 8 in both configurations and bucket 16 was never selected. The baseline reused 3 of 8 re-arrivals in every replicate; the changed configuration reused all 8 (Table~\ref{tab:3-3}).

\begin{table}[!t]
\caption{Changes in reuse and computation with more KV slots}
\label{tab:3-3}
\centering\small
\begin{tabularx}{\columnwidth}{L c c}
\toprule
Metric & 8 KV slots & 16 KV slots \\
\midrule
Re-arrival requests reused & 9/24 & 24/24 \\
\addlinespace[3pt]
Total reused tokens & 9,856 & 26,880 \\
\addlinespace[3pt]
Actual prefill computation of all requests (tokens) & 50,736 & 33,712 \\
\addlinespace[3pt]
Actual prefill computation of re-arrival requests (tokens) & 22,103 & 5,079 \\
\addlinespace[3pt]
Maximum actual batch size & 8 & 8 \\
\addlinespace[3pt]
Total decode steps & 1,964 & 1,994 \\
\bottomrule
\end{tabularx}
\end{table}

Reuse increased by 17,024 tokens, and the prefill computation of re-arrival requests decreased by exactly 17,024 tokens, confirming within the same paired comparison that the survival improvement translates into reduced recomputation. That bucket 16 was unused does not mean the execution path was identical: the counts for buckets 4 and 8 were the same (283 and 654), but bucket 1 changed from 891 to 936 and bucket 2 from 136 to 121, so the total decode step count increased from 1,964 to 1,994. Because the dummy block request occurs when the decode request count is below the execution limit, raising the limit to 16 satisfies this condition even when all eight requests are decoding and may change KV pool events, but the cause of the observed shift was not separated. Because \texttt{batch\_size} changes the number of slots and the execution limit together, we do not interpret this as an intervention on the number of slots alone; the effect on total device time is treated in Section~\ref{sec:5}.

\paragraph{Additional computation cost of reuse failure}
Let C be the reuse amount given by the formula. For the same re-arrival prompt, the difference in prefill work between reusing C tokens and reusing nothing is exactly C tokens. In the sequential experiment the re-arrival prompt was 2,008 tokens in every condition, but only 88 tokens were computed when the KV survived versus all 2,008 when reuse failed, and prefill time increased from about 27 ms to 359 ms, about 13.1-fold. Applying this check to the 168 re-arrivals of the validation measurement that include tool waiting, previous prompts of 800--1,600 tokens give $C$ of 768--1,536 tokens for the 95 requests that reused, and applying the prefill cost model of Section~\ref{sec:3-3} to each request's $L$ and $C$ gives an additional recomputation cost on failure, $\widehat T_{\mathrm{prefill}}(L)-\widehat T_{\mathrm{prefill}}(L-C)$, of about 132 ms for requests with $C=768$ ($L$ of 1,004--1,044) and about 268--270 ms for requests with $C=1{,}536$ ($L$ of 1,653--1,757). In this workload, $L-C$ lies within 65--384 tokens, so the difference from the cost of computing the reusable portion as a separate request, $\widehat T_{\mathrm{prefill}}(C)$, is only 1.0--4.7 ms. For the sequential re-arrival, the model predicts $\widehat T_{\mathrm{prefill}}(2{,}008)-\widehat T_{\mathrm{prefill}}(88)$ of about 338.6 ms versus the measured 332 ms, a difference of about 6.6 ms, but since the 2,008-token point was used in fitting, this is not an independent validation. These per-request costs enter the simulator in Section~\ref{sec:4} as the cost term of reuse failure.

\paragraph{Confirmed range}
The intervention with more slots confirmed that reuse failures are resolved, but because \texttt{batch\_size} also changes the execution limit, this is not the effect of the number of slots alone. The critical request count and the order of lookup versus eviction are observations under the sequential condition. The FIFO rule and the absence of order refresh on re-access are characteristics of this stack confirmed from source and logs, and the structure in which a slot holds one whole sequence is a characteristic of the eager path. In a system that combines fixed KV capacity, sequence-level reservation, and FIFO eviction, the reusable range may be either fully reused or not reused at all.

\subsection{Prefill Interference}\label{sec:3-3}

\paragraph{Exclusive execution and background session delay}
The scheduler of this stack does not place prefill and decode in the same step, so the decode progress of other sessions is delayed while prefill executes. We distinguish the execution cost of prefill (seconds) from the cumulative waiting time incurred by other sessions (session$\cdot$seconds): one prefill that makes several sessions wait is incurred once as cost, whereas the sum of per-session waiting times grows with the number of affected sessions.

\paragraph{Injecting prefill into generating sessions}
While 4 background sessions (300-token prompts, 800-token generation limit) run decode steps, one injected request is sent 3 s after the start. The injected prompt length is 500, 2,000, or 6,000 tokens with a 1-token generation limit. Including a control without injection, each condition was run 3 times, 12 runs in total, with server restarts. All 9 injected requests had 0 cached tokens, so the entire length was subject to prefill. Background delay was observed as the interval between reception times of streaming chunks on the client (response arrival interval); because the chunk token counts were not stored, it is not interpreted as the generation interval between tokens. The interval from transmission to completion of the injected request is the injection interval; for each background session, the median of intervals not overlapping it is the baseline interval and the maximum of overlapping intervals is the delay spike. The injected request's prefill time is the server-reported prefill elapsed time (from first scheduling to first output token, including scheduler and host processing and preemption delay), called server-side prefill elapsed time to distinguish it from pure accelerator time. In all 9 injection runs, the aggregated request count increased by exactly 1 and background sessions continued to receive responses after completion, so the increment of the metric is taken as the injected request's elapsed time.

\paragraph{Increase in response arrival intervals}
Baseline intervals were about 11.6--13.0 ms. Injecting prefill produced long intervals in all 4 background sessions, and the spike intervals of the four sessions overlapped in all 9 runs. Server-side prefill elapsed time and the delay spike increase together with injected length (Table~\ref{tab:3-4}). For the 2,000-token injection, the interval rose from about 12 ms to about 373 ms, showing that the effect of prefill is not confined to that request. In each run, the spread of spike start times across the four sessions was 0.082--0.241 ms and of end times 0.101--0.432 ms, within 1 ms. The preregistered criterion was whether a moment exists at which all four intervals overlap, and it was satisfied. The spike was about 12--15 ms longer than the elapsed time; because the two metrics are measured at different points and the excess was not decomposed, it is not interpreted as the size of a particular execution cost.

\begin{table*}[!t]
\caption{Elapsed time and background session delay by injected prefill length}
\label{tab:3-4}
\centering\small
\begin{tabularx}{\textwidth}{Y Y Y Y}
\toprule
Injected prompt length (tokens) & Server-side prefill elapsed time (ms) & Background session delay spike (ms) & Spike / prefill elapsed time \\
\midrule
500 & 88.5 & 100.9 & 1.140 \\
2,000 & 359.6 & 372.6 & 1.036 \\
6,000 & 1,177.2 & 1,191.7 & 1.012 \\
\bottomrule
\end{tabularx}
\end{table*}

Prefill elapsed time is the median of 3 replicates. The delay spike is the median over the 4 background sessions of the maximum interval in each run, then the median over the three replicates.

In all 36 background-session cases, the spike exceeded the preregistered threshold of 5 times the baseline interval. Intervals exceeding the threshold also appeared in the control runs, all ending within 0.31--0.32 s of the start: immediately after the start, the four background prefills are processed one at a time under exclusive execution, and the decode of an earlier session stops during a later session's prefill, so these are traces of the same exclusive structure independent of injection. The rule was set in advance to conclude only partial confirmation in this case, so the existence decision remains a partial confirmation, while the simultaneity and growth-with-length decisions satisfied their criteria.

\paragraph{Prefill elapsed-time model}
With q the number of tokens actually computed, the server-side prefill elapsed time is approximated by

\begin{equation}
\begin{gathered}
\widehat{T}_{\mathrm{prefill}}(q)=\left\lceil\frac{q}{128}\right\rceil\bigl(0.021206\\
+6.399\times10^{-7}q\bigr)\quad[\mathrm{s}]
\end{gathered}
\end{equation}

with the model value 0 when q is 0. The 128-token unit is the prefix management unit of this stack, adopted as the discretization unit of the cost expression and not interpreted as the number of accelerator invocations. The coefficients were fitted to four observations---500, 2,000, and 6,000 tokens from the injection experiment and 2,008 tokens from Section~\ref{sec:3-2}---with a maximum absolute residual of 2.4 ms; residuals at fitting points do not establish independent accuracy. The fitting range is 500--6,000 tokens, and application below this range is extrapolation. Under the cache-survival condition of Section~\ref{sec:3-2}, the observed time for a residual computation of 88 tokens was 27.2--27.5 ms, whereas the model predicted about 21.3 ms. The model therefore underestimated the observed time by about 6 ms. This error exceeded the maximum residual within the fitting range. The error below the fitting range was evaluated only at this single point. With L the re-arrival prompt length and C the reuse amount on survival, the computation is q = L $-$ C, and the additional cost of reuse failure is $\Delta\widehat{T}_{\mathrm{prefill}}=\widehat{T}_{\mathrm{prefill}}(L)-\widehat{T}_{\mathrm{prefill}}(L-C)$. This is distinguished from the cost of computing the reusable portion $C$ as a separate request, $\widehat{T}_{\mathrm{prefill}}(C)$. Because the model is not linear, the two are not in general equal; within the observed range of this workload ($L-C$ of 65--384 tokens), they differ by 1.0--4.7 ms.

\paragraph{Cumulative waiting time due to interference}
Let $[s_j,e_j]$ be the exclusive execution interval of prefill j and $K_j(t)$ the number of other sessions unable to decode at time t because of it. The cumulative waiting time is

\begin{equation}
W_j=\int_{s_j}^{e_j}K_j(t)\,\mathrm{d}t\quad[\mathrm{session}\cdot\mathrm{s}]
\end{equation}

If the same $K_j$ sessions wait throughout, this simplifies to $W_j=K_jT_j$ with $T_j=e_j-s_j$; using the server-side elapsed time or its model in place of $T_j$ is an approximation. For example, taking 0.3596 s for the 2,000-token injection with four sessions waiting throughout gives about 1.44 session$\cdot$s, which does not mean that each session is delayed by 1.44 s or that the device is occupied for 1.44 s. This session count differs from N: sessions waiting for tools, completed sessions, and requests not yet admitted are not decode-waiting sessions, so the size of interference depends on how much each prefill overlaps other sessions' decode as well as on the number and length of prefills. The experiment fixed the number of background sessions at 4 and did not measure the proportional relation. In cost computation, the prefill time term and the cumulative waiting time are kept separate, and the latter is not added to device time.

\paragraph{Confirmed range}
Exclusive execution of prefill and decode is a characteristic of this stack confirmed from source and logs. The injection experiment observed lengthening of background response intervals accompanying prefill; the existence decision is a partial confirmation, and the simultaneity and growth-with-length criteria are satisfied. Exclusive scheduling is this stack's implementation, but interference in which prefill delays background decode has been reported as generation stalls in GPU serving, with chunked prefill proposed to reduce it \cite{ref14}. This observation is the form that interference takes in a compile-time-static configuration without chunked prefill.

\subsection{Confirmed Range of the Three Mechanisms and Their Coupling through the Arrival Process}\label{sec:3-4}

The three mechanisms each correspond to the execution structure in Section~\ref{sec:2-2}. When tool waiting changes the number of active requests, padding and step count change according to the bucket grid (Section~\ref{sec:3-1}); whether the previous KV is retained until re-arrival is affected by the fixed number of slots and the FIFO eviction order (Section~\ref{sec:3-2}); and while the prefill added by reuse failure executes, the decode of other sessions can be delayed by the exclusive execution structure (Section~\ref{sec:3-3}). The level of confirmation for each mechanism is summarized in Table~\ref{tab:3-5}.

\begin{table*}[!t]
\caption{Confirmed range of the three mechanisms}
\label{tab:3-5}
\centering\small
\begin{tabularx}{\textwidth}{>{\raggedright\arraybackslash}p{2.4cm} L L L}
\toprule
Mechanism & Range confirmed by intervention & Range confirmed by observation & Computed or derived values \\
\midrule
Discrete batch alignment & That the bucket grid is the cause of the padding difference (paired comparison with the same input plan) & Direction of padding change with tool waiting intervals (observation conditional on the measured workload distribution) & Conversion of padding and step count to device time (measured costs applied, cross-checked with channel B) \\
KV cache survival & That reuse failures are resolved in a configuration with more KV slots (\texttt{batch\_size} intervention) & Critical request count, order of lookup and eviction, survival variation in concurrent execution & Exhaustive cross-check of the reuse formula C, difference in recomputation work \\
Prefill interference & Intervention controlling the presence and length of an injected request without reuse (exclusive execution structure retained) & Increase in background session response intervals (existence partially confirmed; simultaneity and growth-with-length criteria satisfied) & Definition of cumulative waiting time and the prefill elapsed-time model \\
\bottomrule
\end{tabularx}
\end{table*}

In actual execution, the three mechanisms operate together in the same arrival process and combine in three ways.

First, the mechanisms follow one another in sequence. KV reuse failure creates additional prefill work (Section~\ref{sec:3-2}), and that prefill can lead, through exclusive execution, to delays of other sessions that are decoding (Section~\ref{sec:3-3}); the size is determined by the reuse lost and the overlap pattern at execution time. The additional computation is bounded by the reuse the existing configuration was losing: in the N=8 slot intervention the re-arrival prefill computation fell by 17,024 tokens, whereas in a separate N=6 paired comparison the baseline already reused 17 of 18 re-arrivals and the total prefill computation fell only from 25,165 to 24,269 tokens (896 tokens), with no difference in two of three replicates. Moreover, with the same recomputation, no background delay arises if no session is waiting for decode, so cumulative waiting time is not determined by the sum of recomputed tokens. The present experiments support each end of this link but do not join them in a single experiment: the injected request was not a re-arrival forced to recompute by eviction, the slot intervention used non-streaming requests and left no response time series, and the server log's 1-second resolution cannot reconstruct the overlap. How much adding slots reduced background delay was therefore not confirmed directly. Even when reuse succeeds, prefill outside the reuse range---the previous generated output and the appended context---remains, so resolving reuse failure does not eliminate prefill interference entirely.

Second, the effects conflict, and the direction depends on conditions. For batch allocation, what matters is which bucket each step's active count maps to; for KV survival, whether the source was evicted before lookup; so the same change in the arrival process does not necessarily move the two metrics in the same direction. In the baseline-grid (\{1,2,4,8\}, N=6) condition of the grid intervention in Section~\ref{sec:3-1}, tool waiting lowered the padding ratio from 0.232 to 0.167 but reduced re-arrival reuse from 18/18 to 14/18 (Table~\ref{tab:3-6}). The direction depends on conditions: with bucket 6 added, the padding ratio at N=6 was 0.057 without tool waiting and 0.098 with it, so the advantage of tool waiting disappeared, and at N=8 on the baseline grid the condition with tool waiting reused 11 of 24, more than 7 of 24 without. The existence of tool waiting must also be distinguished from the variation of waiting times across sessions. A separate N=8 experiment kept the total waiting time fixed and compared equal waiting times for all sessions with session-specific waiting times; reuse was 7 of 24 and 11 of 24, in the predicted direction, but one of three replicates was a tie, so the confirmation criterion was not met and the result is limited to a suggestive observation (that these numbers coincide with the grid-intervention aggregate is a coincidence). Under specific conditions, therefore, tool waiting can reduce padding while worsening KV reuse, and the improvement of one metric alone does not establish that total cost decreased.

\begin{table}[!t]
\caption{Conflict between padding and KV reuse observed at N=6}
\label{tab:3-6}
\centering\small
\begin{tabularx}{\columnwidth}{L Y Y}
\toprule
Metric & Without tool waiting & With tool waiting \\
\midrule
Padding ratio & 0.232 & 0.167 \\
\addlinespace[3pt]
Re-arrival requests reused & 18/18 & 14/18 \\
\bottomrule
\end{tabularx}
\end{table}

Third, execution outcomes feed back into subsequent execution conditions through re-arrival times. Each session sends its next request after receiving the first response and waiting the predetermined time, so under the same waiting-time plan, a configuration change that alters completion times also alters re-arrival times, the number of active requests, and which KV is evicted before lookup. Two observations are consistent with this path. In the N=6 grid-intervention comparison, the condition without tool waiting had 1,288 decode steps on both grids with identical h(n), so its padding change could be decomposed into the selection-rule change alone, whereas with tool waiting the step count changed from 2,001 to 1,970 and h(n) differed. In the N=8 slot-intervention comparison, bucket 16 was unused, yet the counts for buckets 1 and 2 changed and the total rose from 1,964 to 1,994. We do not attribute these distribution changes to re-arrival times alone; in particular, because \texttt{batch\_size} also changes the execution limit, the decode shift in the slot intervention cannot be attributed to reuse improvement alone. What was confirmed is the difference in execution distribution between configurations; feedback through re-arrival is a structural path that can explain it.

This coupling structure means that configuration evaluation must distinguish the inputs to hold fixed from the execution outcomes to recompute. The plan of prompts, generation lengths, and tool waiting times can be shared across configurations, but completion times, re-arrival times, and the variation in active request counts should not be assumed identical, and summing per-mechanism effects obtained from the baseline execution distribution as fixed independent contributions misses the execution process that changes with the configuration. The simulator in Section~\ref{sec:4} accordingly computes request completion and re-arrival, KV eviction and reuse, and prefill and decode execution in time order, and keeps execution cost and cumulative waiting time in different units.

\section{Execution Simulator and Cost Prediction between Configurations}\label{sec:4}

Because the three mechanisms of Section~\ref{sec:3} combine in the same arrival process, per-mechanism effects computed from one run cannot be treated as fixed independent contributions and summed to obtain the cost of another configuration. This section constructs an event-driven simulator that computes each request from arrival to completion in event order, and evaluates the prediction accuracy and residual of cost ratios between configurations, separating reproduction of existing data from prior prediction on new inputs. Explanatory power here means how accurately the cost ratio of a changed configuration to the baseline is predicted; we do not estimate total cost as contributions by mechanism or interpret the residual as the size of any particular unmodeled factor.

\subsection{A Simulator That Reproduces the Request Processing Process}\label{sec:4-1}

The simulator takes a per-session input plan (prompt length, generation length, tool waiting time) and a configuration (bucket set, number of KV slots, execution limit). A request enters a queue on arrival and begins execution depending on the number of running requests and whether KV can be allocated. When a request completes, the next arrival time is computed by adding the session's tool waiting time, so if a configuration change alters completion times, subsequent re-arrival times change accordingly.

KV management is represented by a fixed slot pool and FIFO eviction by allocation order. When admitting a request, the simulator secures the required slot, selects an evictable slot if none is free, and then looks up the remaining KV of the same session to compute the reuse amount. A slot is not made evictable immediately when its request completes; the eviction candidate for the next request is selected first, and only afterward does the completed request's slot become evictable. The reuse amount is computed in 128-token units per Section~\ref{sec:3-2}. With $L_j$ the prompt length of request j and $\widehat C_j$ the predicted reuse, the prefill work is $\widehat q_j=L_j-\widehat C_j$, with $\widehat C_j$ = 0 on failure.

A step that admits a prefill request does not perform decode. In a decode step, the smallest bucket b(n) accommodating the n running requests is selected and one token is generated per request; the first token is treated as produced by prefill, so a request with generation length G needs G$-$1 decode advances. The time of a decode step is

\begin{equation}
t_{\mathrm{step}}(b,n)=f(b)+\alpha+\beta n
\end{equation}

where f(b) is the sum of the median model execution time and the median sampler time from the controlled measurements, $\alpha$ is the fixed additional time per step, and $\beta$ the additional time per running request. The two coefficients were obtained by fitting the residual of the measured step time after subtracting f(b) as a straight line in the number of requests; the source of this residual was not separated (Table~\ref{tab:4-1}). $t_{step}$ is the same value as the per-step component of device time in Section~\ref{sec:2-4}; the simulator advances time by this amount and sums the same amount as cost.

\begin{table}[!t]
\caption{Coefficients of the decode step cost model}
\label{tab:4-1}
\centering\small
\begin{tabular}{lc}
\toprule
Item & Value used in the simulator \\
\midrule
f(1) & 9.870 ms \\
f(2) & 10.420 ms \\
f(4) & 10.825 ms \\
f(8) & 12.970 ms \\
Fixed additional time $\alpha$ & 0.501 ms \\
Additional time per request $\beta$ & 0.0413 ms/request \\
\bottomrule
\end{tabular}
\end{table}

For prefill time, the server-side prefill elapsed-time model of Section~\ref{sec:3-3} is used, with $\widehat{T}_{\mathrm{prefill}}(q)=0$ for $q=0$. The same formula is applied when the remaining work after reuse is below 500 tokens, which is extrapolation outside the coefficient estimation range. We define $\widehat D$ as the sum of the predicted decode and prefill execution times.

\begin{equation}
\widehat D=\sum_{s\in\mathrm{decode}}t_{\mathrm{step}}(b_s,n_s)+\sum_{j\in\mathrm{prefill}}\widehat{T}_{\mathrm{prefill}}(\widehat q_j)
\end{equation}

Intervals in which neither prefill nor decode executes, such as tool waiting, are not included in $\widehat D$. The time other sessions wait for decode because of prefill is computed separately. With $\widehat K_j$ the number of sessions waiting for decode while prefill j executes,

\begin{equation}
\widehat W=\sum_{j\in\mathrm{prefill}}\widehat{T}_{\mathrm{prefill}}(\widehat q_j)\widehat K_j
\end{equation}

$\widehat K_j$ is the number of sessions already decoding when prefill j is admitted and is constant within the interval because the simulator treats a prefill interval as a single event; this is an approximation of the integral definition in Section~\ref{sec:3-3}. $\widehat D$ (seconds) and $\widehat W$ (session$\cdot$seconds) are not summed; the configuration comparisons below evaluate the ratio of $\widehat D$ and do not validate the accuracy of $\widehat W$.

\subsection{Assumptions and Scope of the Simulator}\label{sec:4-2}

The cost coefficients are the earlier measured values and are not readjusted or corrected to match the validation results. Estimating coefficients from measurements, estimating costs for unmeasured buckets, and simplifying execution order are nevertheless assumptions and approximations. Simultaneously arriving requests are assumed to be processed in session index order; the client-side delay from receiving a response to sending the next request is set to 0; and length differences from re-tokenization are not reproduced. Each request is assumed to generate exactly its planned length, which matches the measurements. Prefill cost is a function of q = L $-$ C only, and whether cost is the same for the same q when the reused prefix length differs was not verified. The simulator has queues from the execution limit and the KV allocation constraint but does not separately compute HTTP, host thread, or scheduler-internal time. The dummy block request path of Section~\ref{sec:3-2} and its eviction are not reflected in the base simulator. A simulator that includes this rule had larger prediction errors in 6 of 8 comparisons and smaller errors in none, and its reuse predictions for the baseline configuration moved further from the measurements (Supplementary Material S9). The reason that adding an observed rule worsens prediction was not isolated; an interaction with the eviction order and completion handling rules may be involved. The simulator therefore represents the main paths of the three mechanisms but does not reproduce every event of the actual scheduler and KV management.

\subsection{Comparison Metrics and Validation Design}\label{sec:4-3}

We first checked reproduction of the execution distribution on the existing data referenced in constructing the simulator, and then generated predictions for new input plans not used in configuration selection or simulator development, recording predictions and decision criteria before measurement. The cost validation uses the three configurations of Table~\ref{tab:4-2}, which are also used in the measured evaluation of Section~\ref{sec:5}. The maximum sequence length is 8,192 tokens throughout. The three configurations in each replicate share the same input plan, each condition is measured in three replicates with server restarts, and the confirmatory range is N $\in$ \{6,8\} with N = 10 exploratory.

\begin{table}[!t]
\caption{Configurations used for cost prediction validation and configuration selection evaluation}
\label{tab:4-2}
\centering\small
\begin{tabularx}{\columnwidth}{L c L}
\toprule
Configuration & \texttt{batch\_size} (= KV slots) & Decode bucket set \\
\midrule
Baseline & 8 & \{1, 2, 4, 8\} \\
\addlinespace[3pt]
More KV slots & 16 & \{1, 2, 4, 8, 16\} \\
\addlinespace[3pt]
More KV slots and changed bucket grid & 16 & \{1, 4, 6, 8, 10, 16\} \\
\bottomrule
\end{tabularx}
\end{table}

Comparison with observation uses the two channels of Section~\ref{sec:2-4}. With $c$ a changed configuration, $0$ the baseline, and $k$ a replicate, let $\widehat D_{c,k}$ be the predicted cost and $A_{c,k}$ the channel A aggregate. The cost ratio between configurations and the prediction error are

\begin{equation}
\begin{gathered}
\widehat R_c=\frac{\sum_k\widehat D_{c,k}}{\sum_k\widehat D_{0,k}},\qquad
R^A_c=\frac{\sum_k A_{c,k}}{\sum_k A_{0,k}},\\
e_c=R^A_c-\widehat R_c
\end{gathered}
\end{equation}

These are ratios computed after summing over replicates, not means of per-replicate ratios, and $e_c$ > 0 means the simulator overestimated the savings. Because channel A and the simulator use the same cost coefficients, this comparison mainly evaluates the simulator's step distribution and reuse predictions; the accuracy of the coefficients is checked separately with channel B, which uses none.

In the initial validation, the difference between the two channels' cost ratios had to be 0.02 or less for the prediction decision to proceed; in confirmatory conditions, the absolute prediction error must be 0.03 or less, and the prediction and observation must be on the same side of 1 or both within the preregistered equivalence band [0.98,1.02]. For N=10 only the error is computed, with no pass/fail decision. The initial N=6 measurement did not satisfy the channel agreement requirement and its decision was withheld; before re-measuring with new inputs, the tolerance rule $\tau$ = max(0.02, $r_0$/$B_0$) of Section~\ref{sec:2-4} was registered. This applies only to the new data, and the re-measurement reports results under both requirements.

\subsection{Reproduction of Execution Patterns and Cost Prediction by Configuration}\label{sec:4-4}

Over 80 runs of existing data, the mean absolute error of the padding ratio was 0.007 and the mean absolute relative error of the decode step count was 1.8\%. These are reproduction levels on the data used in constructing the simulator and are not used as prediction errors for new inputs. Table~\ref{tab:4-3} shows the separately performed cost validation.

\begin{table*}[!t]
\caption{Prediction results for cost ratios between configurations}
\label{tab:4-3}
\centering\small
\begin{tabularx}{\textwidth}{L c L Y Y Y L}
\toprule
Measurement & N & Changed configuration & Predicted ratio $\widehat R$ & Observation-based ratio $R^A$ & Error $e$ & Decision \\
\midrule
Initial measurement & 6 & More KV slots & 0.9610 & 0.9793 & +0.0183 & Withheld due to channel disagreement \\
Initial measurement & 6 & More KV slots and changed bucket grid & 0.9466 & 0.9660 & +0.0194 & Withheld due to channel disagreement \\
Initial measurement & 8 & More KV slots & 0.9101 & 0.9175 & +0.0074 & Passed confirmatory criteria \\
Initial measurement & 8 & More KV slots and changed bucket grid & 0.8971 & 0.9028 & +0.0058 & Passed confirmatory criteria \\
Re-measurement with new inputs & 6 & More KV slots & 0.9874 & 0.9941 & +0.0066 & Passed confirmatory criteria \\
Re-measurement with new inputs & 6 & More KV slots and changed bucket grid & 0.9713 & 0.9789 & +0.0076 & Passed confirmatory criteria \\
Initial measurement, exploratory & 10 & More KV slots & 0.9213 & 0.9552 & +0.0339 & Not subject to decision \\
Initial measurement, exploratory & 10 & More KV slots and changed bucket grid & 0.8899 & 0.9264 & +0.0364 & Not subject to decision \\
\bottomrule
\end{tabularx}
\end{table*}

Errors follow unrounded computation and may differ in the last digit from the difference of the displayed ratios. The absolute errors of the four comparisons that passed were 0.0058--0.0076, within the tolerance of 0.03, corresponding to prediction errors of 0.58--0.76 percentage points in savings rate. The two initial N=6 comparisons satisfied the error and direction requirements, but their channel ratio differences of 0.0205 and 0.0236 exceeded the prerequisite, so the withheld decision is retained. In the N=6 re-measurement with new inputs, the revised tolerance was 0.0377 and the actual channel ratio differences were 0.0001 and 0.0063, so the results also pass the original 0.02 requirement and the confirmatory decision does not depend on the relaxation. Passing does not imply a large cost reduction: in the re-measurement, the configuration with more KV slots had an observation-based cost ratio of 0.9941, which was within the preregistered equivalence band $[0.98,1.02]$ and corresponded to a savings rate of 0.59\%. In the exploratory range N=10, both configurations were predicted to improve, but the errors grew to 0.0339 and 0.0364 and are reported as exploratory results by the prior design.

\subsection{Direction and Interpretation of Prediction Residuals}\label{sec:4-5}

The cost ratio error was positive in every comparison in Table~\ref{tab:4-3}: the simulator predicted lower cost ratios for the changed configurations and consistently overestimated savings. This is the direction of the three-replicate sums, not of every individual replicate. Part of the difference is linked to reuse prediction. At N=6 with new inputs, the simulator predicted baseline reuse of 16/18 versus the observed 17/18, with 18/18 for the changed configuration in both, so it expected a larger reducible recomputation than existed; the prefill cost ratio was 0.9308 predicted versus 0.9664 observed, while the decode cost ratios were 1.0000 in both for the more-KV-slots configuration and 0.9802 versus 0.9816 for the changed-grid configuration.

Even so, we do not attribute the error to KV survival prediction or to any single factor. The processing order of simultaneous arrivals, re-tokenization length differences, cost approximation for unmeasured buckets, extrapolation of the prefill cost model, unmodeled eviction events, and execution variation may act together. In particular, the prefill extrapolation does not act symmetrically across configurations, since the more a configuration restores reuse, the larger the share of residual prefill below the fitting range; but because prediction and measurement use the same cost expression, this error enters the residual only when the event composition differs or through the indirect path of time advancement, which was not analyzed separately, so it is not identified as the cause of the residual direction. Errors in opposite directions may also cancel, so a small overall error does not allow the conclusion that each excluded factor is small.

The prediction error e (dimensionless) must also be distinguished from the channel difference r = B $-$ A (seconds). The baseline three-replicate sums are given in Table~\ref{tab:4-4}; the channel difference was about 1.0--1.7 s and contains both the difference in what the channels observe and the error of the cost approximation, so it is not interpreted as the cost of a particular process. Ratios can also partially cancel errors common to numerator and denominator, so a well-matched cost ratio does not mean that absolute execution times match with the same accuracy.

\begin{table*}[!t]
\caption{Time by cost channel for the baseline configuration}
\label{tab:4-4}
\centering\small
\begin{tabular}{lcccc}
\toprule
Measurement & N & Channel A & Channel B & $r=B-A$ \\
\midrule
Initial measurement & 6 & 22.984 s & 24.469 s & 1.485 s \\
Initial measurement & 8 & 32.197 s & 33.441 s & 1.244 s \\
Initial measurement & 10 & 40.040 s & 41.728 s & 1.688 s \\
Re-measurement with new inputs & 6 & 26.110 s & 27.131 s & 1.022 s \\
\bottomrule
\end{tabular}
\end{table*}

To check whether recomputing re-arrival times from completion times contributes to prediction accuracy, we compared the prediction error of the full simulator with that of a simulator in which each session's re-arrival time is fixed at the value observed in the baseline-configuration run and is not recomputed from the completion times of the changed configuration (Table~\ref{tab:4-5}). In 3 of the 4 confirmatory comparisons the full simulator had the smaller absolute error, but in the more-KV-slots configuration on the new N=6 inputs and in the two exploratory N=10 comparisons the fixed-arrival simulator had the smaller error. Over all 8 comparisons each simulator was closer in 4, and the mean absolute errors were similar, 0.0169 and 0.0172. That reflecting the feedback structure of Section~\ref{sec:3-4} in the simulator improves prediction accuracy was therefore not confirmed in these data.

\begin{table*}[!t]
\caption{Prediction error with and without recomputation of re-arrival times}
\label{tab:4-5}
\centering\small
\begin{tabularx}{\textwidth}{L c L Y Y Y Y Y}
\toprule
Measurement & N & Changed configuration & Observation-based ratio $R^A$ & Full simulator ratio & $\lvert e\rvert$ full & Fixed-arrival ratio & $\lvert e\rvert$ fixed \\
\midrule
Initial (withheld) & 6 & More KV slots & 0.9793 & 0.9610 & 0.0183 & 0.9495 & 0.0298 \\
Initial (withheld) & 6 & More KV slots and changed grid & 0.9660 & 0.9466 & 0.0194 & 0.9513 & 0.0146 \\
Initial & 8 & More KV slots & 0.9175 & 0.9101 & 0.0074 & 0.9329 & 0.0154 \\
Initial & 8 & More KV slots and changed grid & 0.9028 & 0.8971 & 0.0058 & 0.9303 & 0.0275 \\
Initial, exploratory & 10 & More KV slots & 0.9552 & 0.9213 & 0.0339 & 0.9386 & 0.0166 \\
Initial, exploratory & 10 & More KV slots and changed grid & 0.9264 & 0.8899 & 0.0364 & 0.9406 & 0.0142 \\
Re-measurement, new inputs & 6 & More KV slots & 0.9941 & 0.9874 & 0.0066 & 0.9937 & 0.0004 \\
Re-measurement, new inputs & 6 & More KV slots and changed grid & 0.9789 & 0.9713 & 0.0076 & 0.9982 & 0.0193 \\
\bottomrule
\end{tabularx}
\end{table*}

The fixed-arrival simulator sets each session's re-arrival time to the value observed in the baseline-configuration run and does not recompute it from the completion times of the changed configuration. The observation-based ratios and the full-simulator ratios and errors are those of Table~\ref{tab:4-3}.

In summary, within the validated range, computing the processing order of requests, the batch composition, and the KV reuse amount together was effective for predicting the cost of each configuration. This is neither an estimate that the three mechanisms explain a certain fraction of total cost nor a validation of cumulative waiting time. Section~\ref{sec:5} uses the simulator to compare candidates while distinguishing the confirmed range from exploratory predictions and taking into account the direction in which the simulator overestimated savings.

\section{Selection of Compile-Time Configurations and Cost Improvement}\label{sec:5}

This section compares configuration candidates with the simulator and presents selection criteria based on gains and costs confirmed by measurement. The central point is that a single compile parameter affects interconnected cost paths. On this stack, \texttt{batch\_size} sets both the number of KV slots and the execution limit, and where the baseline loses substantial reuse, increasing it alone improved KV survival and reduced recomputation prefill, capturing most of the gain of the configuration change. What the intervention confirmed directly is reuse restoration and reduced prefill work; the change in other sessions' waiting time was not measured.

\subsection{Comparing Configuration Candidates with the Simulator}\label{sec:5-1}

The search targets are \texttt{batch\_size} and the decode bucket set, with the language model, numerical precision, maximum sequence length, and number of devices fixed. The \texttt{batch\_size} candidates are 8, 10, 12, and 16; each bucket set includes 1 and the \texttt{batch\_size}, with intermediate buckets chosen from the integers between so that the total is at most 6. The candidate set has 2,077 configurations, and none was excluded by the compile-time limit of 1,800 s. For buckets not measured directly, f(b) is linearly interpolated between the measured points \{1,2,4,8\} and extrapolated along the line through 4 and 8 for buckets above 8, with $\alpha$ and $\beta$ from Section~\ref{sec:4}; interpolated or extrapolated values are not used as reference points for other buckets.

Tool waiting times use the TraceLab distribution, and the per-session plan is applied identically to every configuration. The search uses N $\in$ \{6, 8, 10\}, with 3 seeds and 3 input blocks per N, so the search inputs and the evaluation inputs each consist of 27 plans. The score is the ratio of the candidate's predicted execution cost (decode and prefill) summed over all inputs to the same sum for the baseline; because costs in seconds are summed first, values of N with larger baseline cost carry more weight. The configuration with the smallest score on the search inputs is selected, and the score on the seed-separated evaluation inputs is not used for selection. The selected configuration was \texttt{batch\_size=16} with bucket set \{1, 4, 6, 8, 10, 16\}, with a predicted cost ratio of 0.9066 on the evaluation inputs---a simulator result, distinguished from measured savings. In the re-run search, the top 20 configurations all had \texttt{batch\_size} 16 with predicted ratios within 0.5 percentage points, so their ranking is not interpreted as measured superiority. The selection depended on the concurrency levels included in the search. Restricting the search to the confirmatory concurrency levels $N\in\{6,8\}$ selected the bucket set $\{1,4,5,6,8,16\}$. In this restricted search, the originally selected configuration ranked 11th, and only four configurations were shared between the top 20 candidates from the full and restricted searches. Searching with $N\in\{8,10\}$ selected the original configuration again. By contrast, using the mean of the per-$N$ cost ratios instead of the score based on total cost in seconds did not change the selection for any of the concurrency sets examined (Supplementary Material S8). The configuration reported in this section was thus selected by a search that included $N=10$, which was outside the confirmatory scope. The search identified a set of candidates with similar predicted costs but did not establish the superiority of any particular configuration.

The measurements compare the two changed configurations of Table~\ref{tab:4-2} against the baseline: one increases \texttt{batch\_size} while keeping the existing buckets and adding the top bucket 16, and the other, the selected configuration, also changes the intermediate bucket set. After compiling, we verified that actual bucket selection matched expectation, and performed paired comparisons on new inputs not used in selection, recording predictions and criteria beforehand. This procedure avoids compiling and running every candidate; given the limitations of Section~\ref{sec:4}, the configuration selected by the simulator is not regarded as the global optimum. The execution commands and per-candidate files of the original search are not preserved; a re-run with the published candidate definitions and seeds reproduced the selected configuration and evaluation ratio. To enumerate the same 2,077 configurations with the published code, \texttt{-{}-{}max-{}buckets 5} must be specified (the default 6 includes configurations with 7 buckets).

\subsection{Gains from More KV Slots and a Changed Bucket Grid}\label{sec:5-2}

Table~\ref{tab:5-1} shows the initial N=8 measurement and the N=6 re-measurement with new inputs, both of which passed the confirmatory criteria. Because the two use different input plans, the difference in gain is not interpreted as an effect of N alone.

\begin{table*}[!t]
\caption{Cost savings rates by configuration under confirmatory conditions}
\label{tab:5-1}
\centering\small
\begin{tabularx}{\textwidth}{L L Y Y Y Y}
\toprule
Measurement condition & Changed configuration & Channel A cost ratio & Channel A savings & Channel B cost ratio & Channel B savings \\
\midrule
N=8 initial measurement & More KV slots & 0.9175 & 8.25\% & 0.9183 & 8.17\% \\
\addlinespace[3pt]
N=8 initial measurement & More KV slots and changed bucket grid & 0.9028 & 9.72\% & 0.8993 & 10.07\% \\
\addlinespace[3pt]
N=6 re-measurement with new inputs & More KV slots & 0.9941 & 0.59\% & 0.9940 & 0.60\% \\
\addlinespace[3pt]
N=6 re-measurement with new inputs & More KV slots and changed bucket grid & 0.9789 & 2.11\% & 0.9726 & 2.74\% \\
\bottomrule
\end{tabularx}
\end{table*}

At N=8 the more-KV-slots configuration saved 8.25\% by channel A and the configuration that also changed the grid 9.72\%, a difference of 1.47 percentage points. The path is confirmed in the reuse and prefill computation of Table~\ref{tab:3-3}: the baseline reused 9 of 24 re-arrivals, both changed configurations reused 24 of 24, and the 17,024-token decrease in re-arrival prefill computation matched the increase in reuse exactly. The prefill component of channel A fell to about 67.8\% of the baseline in both changed configurations, whereas the decode component was about 101.3\% for more KV slots and about 99.3\% for the changed grid. In the more-KV-slots configuration, decode cost rose slightly but the prefill decrease offset it; the main savings path was reduced prefill work from restored KV reuse rather than reduced padding. This is not interpreted as confirming reduced delay of other sessions (Section~\ref{sec:3-4}), and 1.47 percentage points is not an independent grid contribution applicable to all configurations.

\subsection{Priority of Configuration Changes by Reuse Loss}\label{sec:5-3}

Adding KV slots does not always give the largest gain. In the N=6 re-measurement, the baseline already reused 17 of 18 re-arrivals; the configuration with more slots reused all 18, but total prefill computation fell by only 896 tokens, with no difference in two of three replicates. The configuration with more KV slots had an observation-based cost ratio of 0.9941, which was within the preregistered equivalence band $[0.98,1.02]$ and corresponded to a savings rate of 0.59\%. Changing the bucket grid as well increased the savings rate to 2.11\%, an additional 1.52 percentage points. Unlike N=8, here the additional gain from changing the grid exceeded the gain from adding slots.

Before selecting a configuration, therefore, one should check how much reuse the baseline is losing. If reuse failures are frequent and the resulting prefill work is large, adding slots has priority; if reuse is mostly retained, the recomputation that further slots can reduce is limited, and a bucket configuration matched to the actual request counts should be examined. The gain cannot be determined from the number of successful reuses alone: each failed request loses a different amount, and for the same token count the prefill cost and the overlap with other requests differ, so the number of additionally computed tokens and the per-step distribution of request counts and buckets should be checked together.

\subsection{Conditions under Which the Gain from Additional KV Slots Is Limited}\label{sec:5-4}

We measured on separate new inputs whether the gain continues as slots are added. Three configurations with \texttt{batch\_size} 16, 24, and 32 keep the common buckets \{1, 4, 6, 8, 10\} and change only the top bucket; the reference is \texttt{batch\_size=8} with \{1, 2, 4, 8\}. Because the comparison from the reference to 16 changes both the number of slots and the intermediate buckets, saturation is evaluated mainly from 16$\rightarrow$24 and 24$\rightarrow$32. Three replicates were run at each N, and all ratios use this experiment's baseline as denominator (Table~\ref{tab:5-2}, channel A). This is a separate experiment with different inputs from Section~\ref{sec:5-2}; the slot-16 bucket set corresponds to the changed-grid configuration of Section~\ref{sec:5-2}, so the results are not chained to it directly.

\begin{table*}[!t]
\caption{Cost and reuse changes with additional KV slots}
\label{tab:5-2}
\centering\small
\begin{tabular}{lcccc}
\toprule
N & Slot 16 cost ratio & Slot 24 cost ratio & Slot 32 cost ratio & Reuse counts: 16 $\rightarrow$ 24 $\rightarrow$ 32 \\
\midrule
6 & 0.9659 & 0.9660 & 0.9663 & 18/18 $\rightarrow$ 18/18 $\rightarrow$ 18/18 \\
8 & 0.9151 & 0.9151 & 0.9161 & 24/24 $\rightarrow$ 24/24 $\rightarrow$ 24/24 \\
10 & 0.8601 & 0.8480 & 0.8482 & 28/30 $\rightarrow$ 30/30 $\rightarrow$ 30/30 \\
\bottomrule
\end{tabular}
\end{table*}

At N=6 and N=8, all re-arrivals were already reused with 16 slots and cost barely changed at 24 and 32. At N=10, two reuse failures remained at 16, all were reused at 24 with the ratio falling from 0.8601 to 0.8480, and 32 brought no further improvement. The preregistered saturation decision classifies an adjacent pair as saturated if the 95\% bootstrap confidence interval of the median ratio over the 9 comparisons (N $\times$ replicate) includes 1 with width 0.04 or less. The medians for 16$\rightarrow$24 and 24$\rightarrow$32 were 0.9999 and 1.0002, satisfying the criterion, but this is an operational classification rule rather than an equivalence test and pools the 9 comparisons, so it does not mean the additional gain at N=10 was absent. The top bucket of the three enlarged configurations was never selected, so enlarging it did not appear as a direct decode cost increase, and where reuse losses were resolved no prefill work remained for additional slots to reduce. This does not present \texttt{batch\_size=16} as a universally optimal value; the point at which to stop adding slots should be judged from the reuse losses on the given inputs and the actual cost change.

\subsection{Compile Cost and Memory Usage}\label{sec:5-5}

A configuration change itself incurs cost. In the measured configurations, one compilation took about 7--8 minutes (Table~\ref{tab:5-3}). This is distinct from per-run cost, and we did not combine compile and execution cost into a single monetary cost or measure the break-even point.

\begin{table}[!t]
\caption{Compile cost of changed configurations}
\label{tab:5-3}
\centering\small
\begin{tabularx}{\columnwidth}{L c Y Y}
\toprule
Configuration & Buckets & Compile time & Compiled artifact size \\
\midrule
Slot 16, existing buckets kept, 16 added & 5 & 407 s & 12.378 GiB \\
\addlinespace[3pt]
Slot 16, changed bucket grid & 6 & 480 s & 13.202 GiB \\
\addlinespace[3pt]
Slot 24, changed bucket grid & 6 & 486 s & 13.259 GiB \\
\addlinespace[3pt]
Slot 32, changed bucket grid & 6 & 474 s & 13.320 GiB \\
\bottomrule
\end{tabularx}
\end{table}

Adding slots also increases the device memory reservation. KV is 144 KiB per token and one slot holds 8,192 tokens, so the KV capacity per slot is 1.125 GiB, or 0.28125 GiB per device across four devices (Table~\ref{tab:5-4}). The slot-32 configuration loaded successfully but, at N=6 and N=8, used more memory without a clear execution cost improvement over slot 16, so fitting on the device is not by itself a reason to choose the largest capacity. The effect of additional memory on co-locating other models or instances was not measured.

\begin{table}[!t]
\caption{Memory usage with additional KV slots}
\label{tab:5-4}
\centering\small
\begin{tabularx}{\columnwidth}{Y Y Y}
\toprule
KV slots & KV capacity per device (arithmetic) & Total memory per device after model loading \\
\midrule
16 & 4.50 GiB & 6.6--7.0 GiB \\
24 & 6.75 GiB & 8.8--9.3 GiB \\
32 & 9.00 GiB & 11.1--11.6 GiB \\
\bottomrule
\end{tabularx}
\end{table}

\subsection{Application Procedure and Scope}\label{sec:5-6}

The configuration selection procedure is as follows. First, diagnose the losses in the baseline: measure reuse failures of re-arrival requests and the additional prefill computation, and check the actual request count and bucket distribution per decode step. Next, for the same input plan, vary the number of KV slots and the bucket set, predict the cost with the simulator, and compare candidates while considering the compile budget and memory usage. Validate the selected configuration on new inputs with predictions and criteria fixed beforehand, confirming actual bucket selection and reuse amounts and cross-checking the two cost channels. Finally, check whether additional slots still reduce execution cost after reuse losses are resolved, and if the additional gain is small, consider a smaller configuration in view of memory and compile cost.

The scope of application is limited to the measured language model and stack, the TraceLab-based tool waiting time distribution, and inputs with one re-arrival per session. For other workloads, the same diagnosis--search--validation procedure should be applied rather than adopting \texttt{batch\_size=16} or a particular bucket set as is; with repeated tool calls, longer contexts, other models, or loads in which the top bucket is actually selected, the relation between reuse loss and execution cost may differ.

\section{Limitations and Conclusion}\label{sec:6}

\subsection{Limitations}\label{sec:6-1}

\paragraph{Workload and validation scope}
This study limits each session to one tool call and did not measure the changes in reuse loss and interference when context accumulates and re-arrivals recur under repeated calls. Synthetic prompts were used in place of real conversations with controlled prompt and generation lengths, so the experiments do not reproduce the content of real conversations or the context composition of individual tools. Configuration search and validation used the TraceLab-based distribution and the controlled experiments on individual mechanisms used separately specified inputs, so the input conditions of each experiment must be distinguished when interpreting results. The confirmatory range of the cost-ratio validation is $N\in\{6,8\}$, with N=10 exploratory. Because the configuration search included all of $N\in\{6,8,10\}$, the search range is wider than the confirmatory range, and excluding N=10 changes the selected configuration (Section~\ref{sec:5-1}). In addition, for loads in which the top bucket of an enlarged configuration is selected continuously, the accuracy of prediction and the effect of configuration selection were not verified.

\paragraph{System and generalization scope}
The results were obtained with the specified language model, software versions, and eager attention path. Other attention implementations use different KV allocation and eviction units, so the observed reuse-termination conditions cannot be applied directly, and discrete bucket selection and exclusive execution of prefill and decode are implementation characteristics of this stack. Applying the findings elsewhere requires checking whether sequence-level KV reservation and FIFO eviction are used together, whether the actual request count is fitted to predefined buckets, and whether prefill delays other sessions' decode, and then measuring the size of each mechanism again. The experiments were limited to a single language model run as one instance; the effect of adding slots on the memory available to co-located models or instances and the cost changes for larger models were not measured.

\paragraph{Measurement, modeling, and decisions}
Neither channel A nor channel B is a direct measurement of the device's pure execution time; the server-side prefill elapsed time may include time other than device computation, and the channel difference r = B $-$ A was not decomposed into particular overheads. The validation in Section~\ref{sec:4} evaluates the accuracy of cost ratios between configurations and does not validate cumulative waiting time, and because the slot intervention stored no response time series, the change in other sessions' waiting time due to restored reuse could not be confirmed directly. The simulator approximates the execution process and costs, with uncertainty in the cost estimation for unmeasured buckets and the extrapolation of the prefill cost expression, so the difference between prediction and measurement is not interpreted as the cost of a particular unmodeled factor, and the possibility that different errors canceled remains. The default tolerance of 0.02 is a predetermined decision criterion; the no-treatment repeated experiment was performed afterward and did not change existing decisions (Supplementary Material S1). The invariance of common bucket costs across recompilation is not fully established, but this concerns the precision of cost conversion and does not affect the padding decomposition of the intervention experiment (Supplementary Material S2). The level from which padding differences due to batch alignment can be distinguished reliably was also not established.

\paragraph{Reproducibility}
The execution commands and per-candidate evaluation files of the original configuration search are not preserved; a re-run with the published search specification and code enumerated the 2,077 candidates and reproduced the selected configuration and evaluation cost ratio. The public repository (\url{https://github.com/dongkyeomjang/escapement}, commit \texttt{3e195aa}) contains the experimental methods, analysis code, decision criteria, and result records (per-table aggregate files under \texttt{results/tables/}) but not all raw data, and validations that depend on separate data such as the TraceLab raw data cannot be completed with the repository alone.

\subsection{Conclusion}\label{sec:6-2}

This paper analyzed, on actual NPU hardware, the costs that arise when the re-arrival of a session after tool waiting interacts with an execution configuration fixed at compile time. The bucket grid intervention with the same input plan confirmed that the bucket configuration is the cause of the padding difference; the KV slot intervention increased the reuse rate and decreased prefill computation by the corresponding amount; and the separate prefill injection experiment lengthened the streaming response intervals of background sessions. These results support the linking structure in which KV survival changes the additional prefill work and that work can slow other sessions' decode, but the reduction in other sessions' delay under the slot intervention was not confirmed directly in a single experiment.

Because the effects of the three mechanisms change together with the arrival and processing order of requests, adding per-mechanism effects as fixed values is insufficient to explain the outcome of a configuration change. A simulator reflecting this, using the previously determined cost coefficients, predicted cost ratios between configurations in the confirmatory range with errors of 0.58--0.76 percentage points. In configuration selection, a single compile parameter was found to affect connected cost paths together: under the N=8 condition with large reuse loss, the configuration with increased \texttt{batch\_size} saved 8.25\% and the configuration that also changed the grid 9.72\%, with the main path being reduced prefill computation from restored reuse, whereas on the N=6 inputs where reuse was mostly retained the gain from adding slots was small. The priority of \texttt{batch\_size} therefore depends on the reuse losses on the current inputs, and once they are resolved, the gain from additional slots must be weighed against the increase in memory usage.

This study presents a method for selecting the number of KV slots and the bucket configuration fixed at compile time to suit the workload. It does not exclude improvements to run-time KV management or scheduling; it is an adjustment whose effect was confirmed by measurement on this stack. Because a padding ratio improvement does not guarantee a reduction in execution cost, resource efficiency cannot be judged from padding alone; actual computation and execution time should be evaluated together, and the waiting time of other sessions and memory usage checked separately according to the objective. Future work should extend validation to repeated tool calls, longer contexts, and loads in which the top bucket of enlarged configurations is used, confirm whether the conditions that produce each mechanism hold on other attention implementations and accelerators, and automate configuration selection that accounts for tool waiting and re-arrival. The analysis code and the experiment and decision records are provided in the public repository cited in Section~\ref{sec:6-1}.

\end{document}


\title{Supplementary Material for\\``Tool Waiting and Re-arrival in Compile-Time-Static LLM Serving: Cost Mechanisms and Configuration Selection''}

\author{Dongkyeom~Jang, In-Nea~Wang, and Junho~Jeong}

\markboth{Supplementary Material}{Jang \MakeLowercase{\textit{et al.}}: Supplementary Material}

\maketitle

\renewcommand{\thesection}{S\arabic{section}}
\renewcommand{\thetable}{S\arabic{table}}

\section{Tolerance Sensitivity and No-Treatment Repeated Measurements}\label{app:S1}

In a sensitivity analysis that varied the default $\tau_0$ in $\tau$ = max($\tau_0$, $r_{BASE}$/$B_{BASE}$) from 0.005 to 0.05, the channel agreement decision was maintained for all comparisons subject to decision. In comparisons where the baseline channel difference ratio $r_{BASE}$/$B_{BASE}$ exceeds $\tau_0$, that ratio determines the tolerance and changing $\tau_0$ does not change it, so this analysis is a sensitivity check on the choice of default, not a validation of the whole tolerance rule.

Whether the tolerance covers the run-to-run variation of the discrepancy between the two channels was checked with no-treatment repeated measurements. With the same input plan and configuration and only the server restarted, 10 runs per N were executed and the distribution of the channel ratio difference $\left|A_i/A_j-B_i/B_j\right|$ between pairs of runs was obtained (Table~\ref{tab:S1}). Single run pairs are the 45 pairs from 10 replicates; 3-run summed pairs are a post hoc aggregate of 2,100 pairs comparing two disjoint sets of 3 runs, which share the underlying runs and are not counted as independent samples. The 95th percentile for single run pairs is wider than $\tau$ at N=6 and within $\tau$ at N=8, but in the 3-run summed unit actually used for decisions, $\tau$ covers the 95th percentile for both N. This measurement was performed after the tolerance was defined and existing decisions were not revised. The distribution describes how differently the two channels quantify the same change in execution; it does not directly represent run-to-run variation of an individual channel or the uncertainty of the savings rate, so it is used only as auxiliary material for reviewing the tolerance and not as a basis for judging whether a small savings effect is distinguishable from execution variation. It also does not correspond to the direction decision criterion at N=7 (S4) or to the range of variation of $\Delta p$.

\begin{table}[H]
\caption{Channel ratio differences in no-treatment repeated measurements}
\label{tab:S1}
\centering\small
\begin{tabularx}{\textwidth}{L Y Y Y Y}
\toprule
Aggregation unit & N & Median & 95th percentile & Maximum \\
\midrule
Single run pair & 6 & 0.0184 & 0.0604 & 0.0656 \\
Single run pair & 8 & 0.0062 & 0.0320 & 0.0343 \\
3-run summed pair & 6 & 0.0134 & 0.0326 & 0.0436 \\
3-run summed pair & 8 & 0.0068 & 0.0167 & 0.0208 \\
\bottomrule
\end{tabularx}
\end{table}

\section{Invariance of Common Bucket Costs across Recompilation}\label{app:S2}

Because the bucket-6 intervention in Section III-A involves recompilation, we checked whether the step costs of the buckets common to both grids (1, 2, 4, 8) were maintained. In the initial decision, the difference for bucket 2 was 0.29 ms, exceeding the preregistered band of 0.03 ms. Since that band did not account for variation between measurement sessions, we rechecked with a repeated design that specifies the variance components. Compared with a recompiled artifact of the same configuration, the differences for all common buckets were within the between-session variation, and the initial 0.29 ms was not reproduced, measuring 0.01 ms. Compared with the artifact with the changed grid, bucket 8 alone exceeded the decision boundary, by one unit of recording resolution (0.01 ms), and this decision is reported as is. The variation range used in the recheck came from sessions within one day and does not cover variation across days. The invariance of common costs is therefore not fully established; this limitation concerns the precision of cost conversion, and the padding decomposition of the intervention experiment, computed from the step sequence and the grid, is not affected by it.

The recheck used 60 lifecycles: three artifacts (A1, the original grid \{1,2,4,8\}; A2, the changed grid \{1,2,4,6,8\}; A3, a recompilation of the same configuration as A1), each restarted and re-measured in 5 sessions for buckets 1, 2, 4, and 8 (Table~\ref{tab:S2}). A session re-measures the same artifact after a restart, whereas a recompilation rebuilds the artifact itself. The variation range was obtained from the absolute difference of median model p50 between session pairs of the same artifact and bucket (10 pairs from 5 sessions, 30 pairs per bucket across three artifacts), and its 95th percentile served as the threshold. Two artifacts are classified as indistinguishable (WITHIN) for a bucket when the difference of their medians is at or below the threshold (Table~\ref{tab:S3}). The recompilation of the same configuration (A1 vs. A3) was WITHIN for all four buckets; the grid change (A1 vs. A2) was OUTSIDE at bucket 8, where the difference of 0.07 ms exceeded the threshold of 0.06 ms by one unit of recording resolution (0.01 ms).

\begin{table}[H]
\caption{Per-artifact, per-bucket measurements in the recompilation variance recheck}
\label{tab:S2}
\centering\footnotesize
\begin{tabularx}{\textwidth}{c c >{\raggedright\arraybackslash}p{4.4cm} Y Y Y Y Y}
\toprule
Bucket & Artifact & Model p50 per session (ms) & Median (ms) & Null pairs & Null median (ms) & Null q95 (ms) & Null max (ms) \\
\midrule
1 & A1 & 9.51 / 9.49 / 9.60 / 9.50 / 9.52 & 9.51 & 30 & 0.04 & 0.22 & 0.24 \\
1 & A2 & 9.53 / 9.49 / 9.49 / 9.60 / 9.47 & 9.49 & 30 & 0.04 & 0.22 & 0.24 \\
1 & A3 & 9.54 / 9.53 / 9.56 / 9.56 / 9.77 & 9.56 & 30 & 0.04 & 0.22 & 0.24 \\
2 & A1 & 10.01 / 10.01 / 10.11 / 10.07 / 10.07 & 10.07 & 30 & 0.06 & 0.14 & 0.18 \\
2 & A2 & 10.12 / 9.94 / 10.06 / 9.99 / 10.09 & 10.06 & 30 & 0.06 & 0.14 & 0.18 \\
2 & A3 & 10.05 / 10.06 / 10.13 / 10.04 / 10.03 & 10.05 & 30 & 0.06 & 0.14 & 0.18 \\
4 & A1 & 10.38 / 10.32 / 10.37 / 10.27 / 10.34 & 10.34 & 30 & 0.05 & 0.37 & 0.40 \\
4 & A2 & 10.32 / 10.32 / 10.47 / 10.31 / 10.32 & 10.32 & 30 & 0.05 & 0.37 & 0.40 \\
4 & A3 & 10.39 / 10.70 / 10.31 / 10.30 / 10.36 & 10.36 & 30 & 0.05 & 0.37 & 0.40 \\
8 & A1 & 12.32 / 12.38 / 12.33 / 12.35 / 12.38 & 12.35 & 30 & 0.03 & 0.06 & 0.06 \\
8 & A2 & 12.40 / 12.42 / 12.41 / 12.44 / 12.46 & 12.42 & 30 & 0.03 & 0.06 & 0.06 \\
8 & A3 & 12.33 / 12.36 / 12.38 / 12.38 / 12.39 & 12.38 & 30 & 0.03 & 0.06 & 0.06 \\
\bottomrule
\end{tabularx}
\end{table}

\begin{table}[H]
\caption{Per-bucket median differences across artifacts and their classification}
\label{tab:S3}
\centering\footnotesize
\begin{tabularx}{\textwidth}{L c Y Y Y Y Y}
\toprule
Comparison & Bucket & Median x (ms) & Median y (ms) & Difference (ms) & q95 (ms) & Classification \\
\midrule
A1 vs. A3 (recompilation) & 1 & 9.51 & 9.56 & 0.05 & 0.22 & WITHIN \\
A1 vs. A3 (recompilation) & 2 & 10.07 & 10.05 & 0.02 & 0.14 & WITHIN \\
A1 vs. A3 (recompilation) & 4 & 10.34 & 10.36 & 0.02 & 0.37 & WITHIN \\
A1 vs. A3 (recompilation) & 8 & 12.35 & 12.38 & 0.03 & 0.06 & WITHIN \\
A1 vs. A2 (grid change) & 1 & 9.51 & 9.49 & 0.02 & 0.22 & WITHIN \\
A1 vs. A2 (grid change) & 2 & 10.07 & 10.06 & 0.01 & 0.14 & WITHIN \\
A1 vs. A2 (grid change) & 4 & 10.34 & 10.32 & 0.02 & 0.37 & WITHIN \\
A1 vs. A2 (grid change) & 8 & 12.35 & 12.42 & 0.07 & 0.06 & OUTSIDE \\
A2 vs. A3 & 1 & 9.49 & 9.56 & 0.07 & 0.22 & WITHIN \\
A2 vs. A3 & 2 & 10.06 & 10.05 & 0.01 & 0.14 & WITHIN \\
A2 vs. A3 & 4 & 10.32 & 10.36 & 0.04 & 0.37 & WITHIN \\
A2 vs. A3 & 8 & 12.42 & 12.38 & 0.04 & 0.06 & WITHIN \\
\bottomrule
\end{tabularx}
\end{table}

Values are the p50 of model execution time with a recording resolution of 0.01 ms. The aggregate file is \texttt{results/tables/S02.md} in the repository.

\section{Attribution of the One Additional Eviction in the Sequential Experiment}\label{app:S3}

In the sequential experiment of Section III-B, the requests that receive slots are the target (1), the background requests (m), and the resumption (1), totaling m+2. If requests were the only consumers of slots, max(0, m$-$6) evictions would occur beyond the pool size of 8, but the observed number was max(0, m$-$5), matching all 21 runs. In the 16 runs with evictions, the count was one higher than the request-allocation computation, and in the remaining 5 runs both values were 0.

The additional eviction is explained by the path in which the scheduler requests a dummy block for padding. "Dummy block" is the name used in the stack's source code; in this configuration it is requested from the same KV pool when the decode request count is greater than 0 and less than the execution limit. This path triggers eviction when no free slot exists but does not allocate the slot to an actual request, so it is not included in the request allocation count. Cross-checking the source path and the logs, there were no preemption-induced evictions, and in each run with evictions exactly one event was found in which the evicted slot was not subsequently reallocated. The logs at that time did not directly show the caller, so this attribution was an explanation combining source and logs; a separate run with added observation that records the call path per event later directly confirmed that the eviction occurs on the dummy block request path. Because the expression was confirmed in replicates of the same condition, we do not claim that it predicts the critical request count for other pool sizes or request compositions. The base simulator in Section IV does not reflect this path or its eviction; the change when it is included is reported in S9.

The direct observation that records the call path was performed in 10 separate runs (excluding one pilot; Table~\ref{tab:S4}). In runs with the sequential-experiment structure (A, 5–8 background requests $\times$ 2) and with the concurrent structure (B, 2 runs), the dummy block was requested exactly once in every partial step in which the decode request count n satisfied 0 < n < 8, and never in steps at the limit (n = 8). There were 14 evictions in total, all on the dummy path (0 on the request-allocation path, 0 by preemption, 0 unclassified), and in each run with evictions exactly one evicted slot was never reallocated. The observation patch was applied only during these runs and then reverted. The server logs are under the repository-relative path \texttt{results/npu/stage2/20260912-{}134732-{}dummy-{}lifecycle/} (the \texttt{results/} directory is not tracked by git), and the aggregate file is \texttt{results/tables/S03.md}.

\begin{table}[H]
\caption{Event counts of the 10 direct-observation runs for the dummy block}
\label{tab:S4}
\centering\footnotesize
\begin{tabularx}{\textwidth}{L Y Y Y Y Y L Y}
\toprule
Run & Decode steps & Partial steps & Dummy requests & Steps at limit & Evictions & Eviction label & Never-reallocated evictions \\
\midrule
A.B5r0 & 49 & 49 & 49 & 0 & 0 & — & 0 \\
A.B5r1 & 49 & 49 & 49 & 0 & 0 & — & 0 \\
A.B6r0 & 56 & 56 & 56 & 0 & 1 & dummy & 1 \\
A.B6r1 & 56 & 56 & 56 & 0 & 1 & dummy & 1 \\
A.B7r0 & 63 & 63 & 63 & 0 & 2 & dummy, dummy & 1 \\
A.B7r1 & 63 & 63 & 63 & 0 & 2 & dummy, dummy & 1 \\
A.B8r0 & 70 & 70 & 70 & 0 & 3 & dummy $\times$3 & 1 \\
A.B8r1 & 70 & 70 & 70 & 0 & 3 & dummy $\times$3 & 1 \\
B.b0 & 535 & 304 & 304 & 231 & 1 & dummy & 1 \\
B.b1 & 535 & 304 & 304 & 231 & 1 & dummy & 1 \\
\bottomrule
\end{tabularx}
\end{table}

\section{Decision Criteria and Per-Block Results at N=7}\label{app:S4}

The metric used to decide the difference between conditions at N=7 is $R=(1-p_A)/(1-p_C)$. Here 1 $-$ p is the fraction of bucket positions occupied by actual requests and does not denote time-based utilization or execution cost. The direction decision uses the preregistered band [0.97, 1.03]; a difference is confirmed when at least 5 of the 6 blocks lie outside the band in the same direction and the overall summed ratio also lies outside it in the same direction. R above the band means the condition with tool waiting has lower padding. Because $\Delta p=(1-p_C)(R-1)$, the fixed-width band for R cannot be applied to $\Delta p$; the main text reports $\Delta p$ but retains the R criterion for decisions.

In the separate validation with new inputs for the bucket set \{1,2,4,8\} and N $\in$ \{3,4,7\}, R is predicted, and the criterion is met when the absolute difference between the predicted ratio summed over 3 new blocks and the measured ratio is 0.05 or less, and the two are on the same side of 1 or both within [0.97,1.03]. This is a criterion for the simulator's prediction accuracy and is distinguished from the between-condition decision here and from the 0.03 cost-ratio error criterion of Section IV.

The padding ratio of each block was computed directly from the stored request counts and bucket sums, not back-calculated from the rounded R (Table~\ref{tab:S5}). Summed ratios are computed after first summing numerators and denominators (Table~\ref{tab:S6}). Of the 6 blocks, 2 are above the band, 4 inside, and 0 below; fewer than 5 lie outside in the same direction and the overall ratio is inside the band, so the preregistered criteria did not confirm a difference. Because the blocks use different input plans, the dispersion of $\Delta p$ includes both input-plan differences and execution variation and is not interpreted as no-treatment variation for the same input.

\begin{table}[H]
\caption{Per-block results at N=7}
\label{tab:S5}
\centering\footnotesize
\begin{tabularx}{\textwidth}{L c Y Y Y Y Y}
\toprule
Measurement batch & Block & $R$ & Direction band & $p_A$ & $p_C$ & $\Delta p$ \\
\midrule
Existing & 1 & 0.9928 & Inside & 0.1687 & 0.1627 & $-$0.0060 \\
Existing & 2 & 1.0229 & Inside & 0.1656 & 0.1843 & +0.0187 \\
Existing & 3 & 1.0481 & Above & 0.1236 & 0.1638 & +0.0402 \\
New & 4 & 0.9922 & Inside & 0.2180 & 0.2118 & $-$0.0062 \\
New & 5 & 1.1032 & Above & 0.1323 & 0.2135 & +0.0811 \\
New & 6 & 1.0140 & Inside & 0.1252 & 0.1373 & +0.0121 \\
\bottomrule
\end{tabularx}
\end{table}

\begin{table}[H]
\caption{Summed results at N=7}
\label{tab:S6}
\centering\small
\begin{tabularx}{\textwidth}{L Y Y}
\toprule
Aggregate & $R$ & $\Delta p$ \\
\midrule
Existing 3 blocks & 1.0220 & +0.0183 \\
New 3 blocks & 1.0322 & +0.0261 \\
All 6 blocks & 1.0273 & +0.0223 \\
\bottomrule
\end{tabularx}
\end{table}

\section{Decomposition of Contributions to the Padding Reduction}\label{app:S5}

The target is the paired comparison at N=6 without tool waiting in which the bucket set changed from \{1,2,4,8\} to \{1,2,4,6,8\}. With h(n) the number of decode steps with n actual requests and b(n) the bucket selected, and h(n) confirmed identical for the two grids, the change in padding ratio can be decomposed into the change in b(n). The rule changes only for n = 5 and n = 6, in both cases from bucket 8 to 6, reducing padding slots by 2 per step. This premise was confirmed for this comparison and is not generalized to other input plans or to conditions with tool waiting.

The first method is the share $D_n/\sum_n D_n$ of the padding slots reduced at each n, $D_n=h(n)\bigl[b_{\mathrm{old}}(n)-b_{\mathrm{new}}(n)\bigr]$; it decomposes the slots removed from the numerator of the padding ratio and does not reflect the change in the denominator. The second is a two-factor Shapley decomposition that allocates the drop of the padding ratio $p=\sum_n h(n)[b(n)-n]/\sum_n h(n)b(n)$, whose numerator and denominator both change, by averaging over the two application orders. With S$\subseteq$\{5,6\} the set of request counts to which the change is applied and $p_S$ the ratio under the corresponding rule,

\begin{equation}
\begin{aligned}
\phi_5&=\frac12\left[(p_{\varnothing}-p_{\{5\}})+(p_{\{6\}}-p_{\{5,6\}})\right]\\
\phi_6&=\frac12\left[(p_{\varnothing}-p_{\{6\}})+(p_{\{5\}}-p_{\{5,6\}})\right]
\end{aligned}
\end{equation}

and the two contributions sum to the total drop $p_{\varnothing}-p_{\{5,6\}}$. This is an allocation on the same data and not a validation independent of the first method.

Over the 1,288 decode steps summed across three replicates, the padding ratio decreased from about 0.2324 to 0.0574, with an unrounded drop of 0.1749263916 (Table~\ref{tab:S7}). The total number of padding slots fell from 1,525 to 307 and the sum of bucket widths from 6,563 to 5,345; the difference between the sum of the Shapley contributions and the total drop was 0 in re-aggregation. Since the per-step reduction is 2 for both n=5 and n=6, the difference in contribution arises from the step counts of 115 and 494. These are contributions to the padding ratio on a decode-step basis; because time cost differs by bucket, the finding that 80.9\% of the drop occurred at n=6 is not interpreted as 80.9\% of the device time savings.

\begin{table}[H]
\caption{Contribution to padding reduction by active request count}
\label{tab:S7}
\centering\footnotesize
\begin{tabularx}{\textwidth}{c c Y Y Y Y Y}
\toprule
Actual requests $n$ & $h(n)$ & Bucket change & Padding slots reduced & Share of slot reduction & Contribution to $p$ drop & Share of drop \\
\midrule
5 & 115 & 8 $\rightarrow$ 6 & 230 & 18.9\% & 0.0333824 & 19.1\% \\
6 & 494 & 8 $\rightarrow$ 6 & 988 & 81.1\% & 0.1415440 & 80.9\% \\
Other & 679 & No change & 0 & 0\% & 0 & 0\% \\
Total & 1,288 & — & 1,218 & 100\% & 0.1749264 & 100\% \\
\bottomrule
\end{tabularx}
\end{table}

\section{Per-Replicate Decomposition of the Validation Comparisons and Savings Rates}\label{app:S6}

The cost ratios in Tables IX and XII of the main text are computed from costs summed across three replicates: the cost of each configuration is first summed over the three replicates and then divided by the summed cost of the baseline configuration. Table~\ref{tab:S8} decomposes those summed costs by replicate. The summed row is the preregistered decision unit; the replicate rows show its composition and are not uncertainty intervals. The no-treatment repeated distribution for the same input is in S1. The savings rate X is 1 $-$ ratio. The aggregate file is \texttt{results/tables/S06a.md}.

\begin{table}[H]
\caption{Per-replicate decomposition of the validation comparisons and savings rates}
\label{tab:S8}
\centering\footnotesize
\begin{tabularx}{\textwidth}{L c L c Y Y Y Y Y Y Y}
\toprule
Measurement & N & Changed configuration & Replicate & Baseline A (s) & Changed A (s) & A ratio & B ratio & Channel diff. & X(A) & X(B) \\
\midrule
Initial & 6 & More KV slots & 1 & 7.696 & 7.694 & 0.9998 & 0.9683 & 0.0315 & +0.02\% & +3.17\% \\
Initial & 6 & More KV slots & 2 & 6.785 & 6.563 & 0.9672 & 0.9699 & 0.0027 & +3.28\% & +3.01\% \\
Initial & 6 & More KV slots & 3 & 8.503 & 8.252 & 0.9705 & 0.9419 & 0.0286 & +2.95\% & +5.81\% \\
Initial & 6 & More KV slots & Sum & 22.984 & 22.508 & 0.9793 & 0.9588 & 0.0205 & +2.07\% & +4.12\% \\
Initial & 6 & Slots and grid & 1 & 7.696 & 7.528 & 0.9782 & 0.9515 & 0.0267 & +2.18\% & +4.85\% \\
Initial & 6 & Slots and grid & 2 & 6.785 & 6.542 & 0.9642 & 0.9573 & 0.0069 & +3.58\% & +4.27\% \\
Initial & 6 & Slots and grid & 3 & 8.503 & 8.131 & 0.9562 & 0.9228 & 0.0334 & +4.38\% & +7.72\% \\
Initial & 6 & Slots and grid & Sum & 22.984 & 22.201 & 0.9660 & 0.9424 & 0.0236 & +3.40\% & +5.76\% \\
Initial & 8 & More KV slots & 1 & 13.441 & 12.697 & 0.9446 & 0.9334 & 0.0112 & +5.54\% & +6.66\% \\
Initial & 8 & More KV slots & 2 & 10.920 & 10.009 & 0.9166 & 0.9256 & 0.0090 & +8.34\% & +7.44\% \\
Initial & 8 & More KV slots & 3 & 7.836 & 6.835 & 0.8723 & 0.8817 & 0.0094 & +12.77\% & +11.83\% \\
Initial & 8 & More KV slots & Sum & 32.197 & 29.541 & 0.9175 & 0.9183 & 0.0008 & +8.25\% & +8.17\% \\
Initial & 8 & Slots and grid & 1 & 13.441 & 12.446 & 0.9259 & 0.9098 & 0.0162 & +7.41\% & +9.02\% \\
Initial & 8 & Slots and grid & 2 & 10.920 & 9.876 & 0.9044 & 0.9095 & 0.0052 & +9.56\% & +9.05\% \\
Initial & 8 & Slots and grid & 3 & 7.836 & 6.747 & 0.8611 & 0.8669 & 0.0058 & +13.89\% & +13.31\% \\
Initial & 8 & Slots and grid & Sum & 32.197 & 29.069 & 0.9028 & 0.8993 & 0.0035 & +9.72\% & +10.07\% \\
Initial, exploratory & 10 & More KV slots & 1 & 12.949 & 12.217 & 0.9434 & 0.9487 & 0.0053 & +5.66\% & +5.13\% \\
Initial, exploratory & 10 & More KV slots & 2 & 13.575 & 12.979 & 0.9560 & 0.9509 & 0.0051 & +4.40\% & +4.91\% \\
Initial, exploratory & 10 & More KV slots & 3 & 13.516 & 13.050 & 0.9655 & 0.9656 & 0.0000 & +3.45\% & +3.44\% \\
Initial, exploratory & 10 & More KV slots & Sum & 40.040 & 38.245 & 0.9552 & 0.9551 & 0.0000 & +4.48\% & +4.49\% \\
Initial, exploratory & 10 & Slots and grid & 1 & 12.949 & 11.438 & 0.8833 & 0.8804 & 0.0029 & +11.67\% & +11.96\% \\
Initial, exploratory & 10 & Slots and grid & 2 & 13.575 & 12.911 & 0.9511 & 0.9542 & 0.0031 & +4.89\% & +4.58\% \\
Initial, exploratory & 10 & Slots and grid & 3 & 13.516 & 12.742 & 0.9428 & 0.9500 & 0.0072 & +5.72\% & +5.00\% \\
Initial, exploratory & 10 & Slots and grid & Sum & 40.040 & 37.091 & 0.9264 & 0.9287 & 0.0023 & +7.36\% & +7.13\% \\
New inputs & 6 & More KV slots & 1 & 9.790 & 9.614 & 0.9821 & 0.9838 & 0.0017 & +1.79\% & +1.62\% \\
New inputs & 6 & More KV slots & 2 & 8.396 & 8.417 & 1.0024 & 0.9991 & 0.0033 & $-$0.24\% & +0.09\% \\
New inputs & 6 & More KV slots & 3 & 7.923 & 7.923 & 1.0000 & 1.0011 & 0.0011 & +0.00\% & $-$0.11\% \\
New inputs & 6 & More KV slots & Sum & 26.110 & 25.954 & 0.9941 & 0.9940 & 0.0001 & +0.59\% & +0.60\% \\
New inputs & 6 & Slots and grid & 1 & 9.790 & 9.561 & 0.9766 & 0.9687 & 0.0080 & +2.34\% & +3.13\% \\
New inputs & 6 & Slots and grid & 2 & 8.396 & 8.258 & 0.9835 & 0.9748 & 0.0087 & +1.65\% & +2.52\% \\
New inputs & 6 & Slots and grid & 3 & 7.923 & 7.739 & 0.9767 & 0.9752 & 0.0015 & +2.33\% & +2.48\% \\
New inputs & 6 & Slots and grid & Sum & 26.110 & 25.558 & 0.9789 & 0.9726 & 0.0063 & +2.11\% & +2.74\% \\
\bottomrule
\end{tabularx}
\end{table}

\section{N $\times$ Replicate Decomposition of the Additional-Slot Saturation}\label{app:S7}

Table~\ref{tab:S9} decomposes the cost ratios of Table XIII by N and replicate. The reference for each row is the \texttt{batch\_size=8} configuration of the same (N, replicate). That the selection share of the top bucket (16, 24, 32) is 0\% in every row is the observation in Section V-D. The aggregate file is \texttt{results/tables/S06b.md}.

\begin{table}[H]
\caption{N $\times$ replicate decomposition of the additional-slot saturation}
\label{tab:S9}
\centering\footnotesize
\begin{tabularx}{\textwidth}{c c Y Y Y Y Y Y L}
\toprule
N & Replicate & Slot 16 A ratio & Slot 16 B ratio & Slot 24 A ratio & Slot 24 B ratio & Slot 32 A ratio & Slot 32 B ratio & Reuse (16 / 24 / 32) \\
\midrule
6 & 1 & 0.9633 & 0.9612 & 0.9636 & 0.9693 & 0.9632 & 0.9571 & 6/6 / 6/6 / 6/6 \\
6 & 2 & 0.9847 & 0.9739 & 0.9845 & 0.9857 & 0.9848 & 0.9733 & 6/6 / 6/6 / 6/6 \\
6 & 3 & 0.9452 & 0.9492 & 0.9452 & 0.9479 & 0.9466 & 0.9447 & 6/6 / 6/6 / 6/6 \\
8 & 1 & 0.9062 & 0.8998 & 0.9061 & 0.9009 & 0.9086 & 0.9084 & 8/8 / 8/8 / 8/8 \\
8 & 2 & 0.9548 & 0.9588 & 0.9548 & 0.9563 & 0.9548 & 0.9567 & 8/8 / 8/8 / 8/8 \\
8 & 3 & 0.8819 & 0.8819 & 0.8819 & 0.8835 & 0.8828 & 0.8803 & 8/8 / 8/8 / 8/8 \\
10 & 1 & 0.8211 & 0.8340 & 0.8208 & 0.8302 & 0.8208 & 0.8297 & 10/10 / 10/10 / 10/10 \\
10 & 2 & 0.8428 & 0.8269 & 0.8261 & 0.8093 & 0.8260 & 0.8081 & 9/10 / 10/10 / 10/10 \\
10 & 3 & 0.9072 & 0.9115 & 0.8898 & 0.8940 & 0.8905 & 0.8946 & 9/10 / 10/10 / 10/10 \\
\bottomrule
\end{tabularx}
\end{table}

\section{Sensitivity of the Configuration Selection to the Set of N and the Score Definition}\label{app:S8}

The configuration search of Section V-A was re-run with different sets of concurrency levels and score definitions (Table~\ref{tab:S10}). The candidate space is the same 2,077 configurations in every condition; the set of N changes only the score. The score was computed both as the recorded rule, the ratio of summed seconds (sum-seconds), and as the unweighted mean of per-N cost ratios (per-n); both values are always reported, and the ranking follows the designated one. The selected configuration and the evaluation ratio of 0.9066 in the reference condition (sum-seconds, N $\in$ \{6, 8, 10\}) match the preregistered record. Table~\ref{tab:S11} compares the top-20 set of each condition with that of the reference condition. The aggregate file is \texttt{results/tables/B01.md}.

\begin{table}[H]
\caption{Selected configuration by set of N and score definition}
\label{tab:S10}
\centering\footnotesize
\begin{tabularx}{\textwidth}{L L L c Y Y Y Y Y Y}
\toprule
Score & Set of N & Selected configuration & batch\_size & Search sum ratio & Search per-N mean & Eval. sum ratio & Eval. per-N mean & Matches record & Rank of recorded config. \\
\midrule
sum-seconds & 6, 8 & \{1, 4, 5, 6, 8, 16\} & 16 & 0.9362 & 0.9383 & 0.9237 & 0.9247 & No & 11 \\
sum-seconds & 6, 8, 10 & \{1, 4, 6, 8, 10, 16\} & 16 & 0.9209 & 0.9254 & 0.9066 & 0.9109 & Yes & 1 \\
sum-seconds & 8, 10 & \{1, 4, 6, 8, 10, 16\} & 16 & 0.9089 & 0.9099 & 0.8870 & 0.8884 & Yes & 1 \\
per-n & 6, 8 & \{1, 4, 5, 6, 8, 16\} & 16 & 0.9362 & 0.9383 & 0.9237 & 0.9247 & No & 12 \\
per-n & 6, 8, 10 & \{1, 4, 6, 8, 10, 16\} & 16 & 0.9209 & 0.9254 & 0.9066 & 0.9109 & Yes & 1 \\
per-n & 8, 10 & \{1, 4, 6, 8, 10, 16\} & 16 & 0.9089 & 0.9099 & 0.8870 & 0.8884 & Yes & 1 \\
\bottomrule
\end{tabularx}
\end{table}

\begin{table}[H]
\caption{Top-20 sets by condition compared with the reference condition}
\label{tab:S11}
\centering\small
\begin{tabularx}{\textwidth}{L L Y Y Y}
\toprule
Score & Set of N & Common with reference & Reference only & This condition only \\
\midrule
sum-seconds & 6, 8 & 4 & 16 & 16 \\
sum-seconds & 6, 8, 10 & 20 & 0 & 0 \\
sum-seconds & 8, 10 & 15 & 5 & 5 \\
per-n & 6, 8 & 4 & 16 & 16 \\
per-n & 6, 8, 10 & 20 & 0 & 0 \\
per-n & 8, 10 & 14 & 6 & 6 \\
\bottomrule
\end{tabularx}
\end{table}

\section{Change When the Dummy Block Rule Is Included in the Simulator}\label{app:S9}

The rule observed in S3 (when the decode request count n satisfies 0 < n < execution limit, one additional slot is occupied) was added to the simulator of Section IV as a switch, and the 8 comparisons of Table IX were recomputed with the default setting (not included) and with the rule included (Table~\ref{tab:S12}). The execution limit is \texttt{batch\_size}. The error $\lvert e\rvert$ is the absolute difference between the simulator ratio and the observation-based ratio; the baseline rows have a ratio of 1 by definition. With the rule included, the error grew in 6 of the 8 comparisons and shrank in none; the effect concentrated on the baseline configuration with \texttt{batch\_size} 8, whose simulated reuse moved further from the measurements. Re-running the configuration search of Section V-A with the rule included leaves the selected configuration and the top-20 set unchanged and lowers only the evaluation sum ratio from 0.9066 to 0.8798. The 8 default-setting predictions match the preregistered predictions to four decimal places. The aggregate file is \texttt{results/tables/S08.md}.

\begin{table}[H]
\caption{Predicted ratios and simulated reuse before and after including the dummy block rule}
\label{tab:S12}
\centering\footnotesize
\begin{tabularx}{\textwidth}{L c L Y Y Y Y Y L Y}
\toprule
Measurement & N & Configuration & Observed ratio & Not included: ratio & Not included: $\lvert e\rvert$ & Included: ratio & Included: $\lvert e\rvert$ & Simulated reuse (not incl.$\rightarrow$incl.) & Measured reuse \\
\midrule
Initial & 6 & Baseline & 1.0000 & 1.0000 & — & 1.0000 & — & 14$\rightarrow$13 / 18 & 15/18 \\
Initial & 6 & More KV slots & 0.9793 & 0.9610 & 0.0183 & 0.9576 & 0.0217 & 18$\rightarrow$18 / 18 & 18/18 \\
Initial & 6 & Slots and grid & 0.9660 & 0.9466 & 0.0194 & 0.9432 & 0.0228 & 18$\rightarrow$18 / 18 & 18/18 \\
Initial & 8 & Baseline & 1.0000 & 1.0000 & — & 1.0000 & — & 9$\rightarrow$6 / 24 & 9/24 \\
Initial & 8 & More KV slots & 0.9175 & 0.9101 & 0.0074 & 0.8845 & 0.0330 & 24$\rightarrow$24 / 24 & 24/24 \\
Initial & 8 & Slots and grid & 0.9028 & 0.8971 & 0.0058 & 0.8718 & 0.0310 & 24$\rightarrow$24 / 24 & 24/24 \\
Initial, exploratory & 10 & Baseline & 1.0000 & 1.0000 & — & 1.0000 & — & 7$\rightarrow$6 / 30 & 9/30 \\
Initial, exploratory & 10 & More KV slots & 0.9552 & 0.9213 & 0.0339 & 0.9037 & 0.0515 & 29$\rightarrow$29 / 30 & 27/30 \\
Initial, exploratory & 10 & Slots and grid & 0.9264 & 0.8899 & 0.0364 & 0.8729 & 0.0534 & 29$\rightarrow$29 / 30 & 27/30 \\
New inputs & 6 & Baseline & 1.0000 & 1.0000 & — & 1.0000 & — & 16$\rightarrow$16 / 18 & 17/18 \\
New inputs & 6 & More KV slots & 0.9941 & 0.9874 & 0.0066 & 0.9874 & 0.0066 & 18$\rightarrow$18 / 18 & 18/18 \\
New inputs & 6 & Slots and grid & 0.9789 & 0.9713 & 0.0076 & 0.9713 & 0.0076 & 18$\rightarrow$18 / 18 & 18/18 \\
\bottomrule
\end{tabularx}
\end{table}